\documentclass[twocolumn]{aastex63}
\hypersetup{linkcolor=blue,citecolor=blue,filecolor=cyan,urlcolor=cyan}

\graphicspath{{./}{figures/}}

\accepted{18 September 2019}
\submitjournal{The Astronomical Journal}

\shorttitle{The role of cluster mass in MPs of GCs}
\shortauthors{Lagioia et al.}

\newcommand{\ngc}{NGC\,}
\newcommand{\hst}{\textit{HST}}

\newcommand{\feh}{\rm [Fe/H]}

\newcommand{\teff}{$T_{\rm eff}$}

\defcitealias{dotter10}{D10}
\defcitealias{marin-franch09}{M09}
\defcitealias{milone17}{M17}
\defcitealias{vandenberg13}{V13}

\begin{document}

\title{The role of cluster mass in the multiple populations of Galactic and extragalactic globular clusters} 

\correspondingauthor{Edoardo Lagioia}
\email{edoardo.lagioia@unipd.it}

\author[0000-0003-1713-0082]{Edoardo P. Lagioia} 
\affiliation{Dipartimento di Fisica e Astronomia ``Galileo Galilei'', Universit\`{a} di Padova, Vicolo dell'Osservatorio 3, I-35122, Padua, Italy}

\author[0000-0001-7506-930X]{Antonino P. Milone}
\affiliation{Dipartimento di Fisica e Astronomia ``Galileo Galilei'', Universit\`{a} di Padova, Vicolo dell'Osservatorio 3, I-35122, Padua, Italy}

\author[0000-0002-1276-5487]{Anna F. Marino}
\affiliation{Dipartimento di Fisica e Astronomia ``Galileo Galilei'', Universit\`{a} di Padova, Vicolo dell'Osservatorio 3, I-35122, Padua, Italy}
\affiliation{Centro di Ateneo di Studi e Attivit\`{a} Spaziali ``Giuseppe Colombo'' - CISAS, via Venezia 15, I-35131, Padua, Italy}

\author[0000-0002-7690-7683]{Giacomo Cordoni}
\affiliation{Dipartimento di Fisica e Astronomia ``Galileo Galilei'', Universit\`{a} di Padova, Vicolo dell'Osservatorio 3, I-35122, Padua, Italy}

\author[0000-0002-1128-098X]{Marco Tailo}
\affiliation{Dipartimento di Fisica e Astronomia ``Galileo Galilei'', Universit\`{a} di Padova, Vicolo dell'Osservatorio 3, I-35122, Padua, Italy}

\begin{abstract}
	Studies based on \textit{Hubble Space Telescope} (\hst) photometry in
	F275W, F336W, and F438W have shown that the incidence and complexity of
	multiple populations (MPs) in Globular Clusters (GCs) depend on cluster
	mass. This result, however, is based on nearby, low-reddening Galactic
	GCs, for which we can obtain accurate F275W photometry.

	In this work we exploit \hst\ photometry in F336W, F438W, and F814W to
	characterize MPs in 68 Galactic and extragalactic GCs by using the
	pseudo-color $C_{F336W,F438W,F814W}$, which is more efficient in terms
	of observation time than the pseudo-color $C_{F275W,F336W,F438W}$
	adopted in previous works.

	We first analyzed the Galactic GCs alone and find that the RGB color
	width strongly correlates with \feh. After removing the dependence
	from metallicity, we obtain a strong correlation with cluster mass,
	thus confirming previous findings.

	We then exploited the RGB width to compare MPs in Galactic and
	extragalactic GCs.  Similarly to Galactic GCs, the RGB width of
	Magellanic Cloud GCs correlates with clusters mass, after removing the
	dependence from metallicity. This fact demonstrates that cluster mass is
	the main factor affecting the properties of MPs.

	Magellanic Cloud clusters exhibit, on average, narrower RGBs than
	Galactic GCs with similar mass and metallicity. We suggest that this
	difference is a signature of stellar mass loss in Galactic GCs.  As an
	alternative, the mass-RGB-width relation would depend on the host
	galaxy.

	Finally, we used ground-based photometry to investigate Terzan\,7
	through the $C_{U,B,I}$ pseudo-color, ground-based analogous of
	$C_{F336W,F438W,F814W}$, and find that this cluster is consistent with
	a simple population.  
\end{abstract}


\section{Introduction} \label{sec:intro}
Growing evidence of the unique properties of globular clusters (GCs) has
accumulated in the last two decades. The discovery of different groups of
stars, with well-defined abundance patterns of specific chemical elements,
namely C, N, O, Na, and in some cases also Mg and Al, known as `multiple
populations' (MPs), turned out to be a peculiar characteristic of almost all
GCs in our Galaxy \citep[e.g.][and references
therein]{kraft94,gratton12a,marino19a}. 

A general feature of GCs is the presence of two main stellar populations, one
with chemical composition typical of halo stars of the same metallicity,
usually referred to as primordial population or `1G' stars, and another with
abundance of N and Na higher than that of 1G stars, referred to as enriched
population or `2G' stars. Furthermore, 2G stars are typically also enriched in
helium \citep[e.g.][]{lagioia18,milone18b}.  Both 2G and, to a lesser extent,
1G stars contain, however, a varying number of subgroups of stars, each one
with different chemical content \citep[e.g.][]{milone15a,milone15c,marino19a},
while the ratio of 1G to 2G stars is not constant in all GCs but ranges from
about 10\% to 60\% \citep{milone17}. The composite phenomenology of MPs
suggests that peculiar processes of star formation,
chemical evolution and dynamical interactions are at work in the internal
environment of star clusters \cite[e.g.][]{decressin07,dantona16,gieles18}. 

A comprehensive framework for the interpretation of the multifaceted
characteristics of MPs requires a common metrics that has the potential of
quantifying the extension of chemical spread in a given cluster. To this
purpose, an effective solution is provided by the color extension, or width,
of the red giant branch (RGB) in color-magnitude diagrams (CMDs) obtained
with proper combinations of ultraviolet (UV) and optical bands.  

A systematic measurement of the RGB width in the color $m_{F275W}-m_{F814W}$
and in the pseudo-color $(m_{F275W}-m_{F336W})-(m_{F336W}-m_{F438W})$ has been
recently performed by \citet{milone17} for the 57 GCs observed for the
\textit{Hubble Space Telescope} (\hst) UV Legacy Survey of Galactic GCs
\citep[see][and references therein]{piotto15}. An important result of this work
concerns the statistical analysis of the correlation between the global
parameters of GCs and the RGB width. It revealed that, after taking into
account the contribution of the metallicity to the observed spread, the RGB
width appears to be strongly correlated with the cluster mass, thus suggesting
a major role for this parameter in the determination of the observed MP
properties. 

The sample of clusters studied by Milone and collaborators comprises only
Galactic GCs older than $\sim 12$\,Gyr \citep[e.g.][]{dotter10}. It is
challenging, therefore, to infer strong constraints on the effect of cluster
age and on the role of the host galaxy on the onset of MPs from their sample
only. The recent discovery of the presence of MPs in Magellanic Clouds (MCs)
\citep[e.g.][]{lagioia19, martocchia18a, niederhofer17} and Fornax GCs
\citep[e.g.][]{larsen14a,larsen12a} indicates that the MP phenomenon is not
limited to our Galaxy but rather a common characteristic of GCs.  In this
regard, the comparison of the observational features of Galactic and
extragalactic GCs presents us the opportunity to test the role played by the
external environment in the determination of the MP properties.

As seen above, the results obtained by \citet{milone17} rely on F275W band
data, which are sensitive to the oxygen abundance through the spectral
absorption of the OH molecule. This benefit, however, is counterbalanced by the
difficulty to obtain F275W images with sufficiently high S/N for the most
absorbed or the most distant clusters.

The pseudo-color $C_{U,B,I} = (U-B)-(B-I)$ \citep{monelli13} offers a valuable
alternative to investigate MPs in GCs. As demonstrated by \citet{marino08} the
broad-band filters $U$ and $B$ are ideal to photometrically disentangle carbon
and nitrogen variations in RGB stars, while the large baseline offered by the
optical color $B-I$, makes the $C_{U,B,I}$ also sensitive to the helium content
of stars through the effective temperature \citep[\teff;][]{milone12a}. For this
reason, the spread of the RGB sequence in this pseudo-color is proportional to
the variation of chemical content among the underlying MPs.  

With the aim of comparing the MP properties of a sample of 68 Galactic and
extragalactic GCs, we exploit the RGB width in the pseudo-color
$C_{F336W,F438W,F814W} = (m_{F336W}-m_{F438W})-(m_{F438W}-m_{F814W})$  to
investigate the relationships between MPs and a series of morphological,
structural and physical GC parameters. The choice of the three \hst\ bands
F336W, F438W and F814W, which are analogous to the Johnson-Cousins U, B and I
bands, is justified by three empirical reasons: i) the aforementioned
sensitivity of these bands to the C, N, O and He content of GC RGB stars; ii)
the availability of a large number of archival observations in these three
bands, in the archives of both \hst\ and ground-based telescopes; iii) the
foreclosure of far-UV observations in the post-\hst\ era.  Moreover, for a
fixed signal-to-noise ratio, observations in F336W, F438W, and F814W require
much shorted \hst\ time than F275W, F336W and F438W. As a
consequence, this filters make it possible to investigate the MP phenomenon in
distant clusters. In addition, we exploit $U, B, I$ ground-based photometry to
investigate the GC Terzan\,7.

This paper is organized as follows: in Section~\ref{sec:obs} we describe the
dataset and the techniques for the analysis of the photometric data; in
Section~\ref{sec:rgbw} we define the RGB width and describe the methodology to
measure this quantity and the corresponding error for all the GCs in our
database; in Section~\ref{sec:corr} we perform a statistical analysis to test
the correlation between the RGB width and a set of GC global parameters; in
Section~\ref{sec:ncorr}, we perform a similar analysis after removing the
effect of metallicity on the RGB width; in Section~\ref{sec:general}, by means
of synthetic spectra analysis, we study the behavior of the theoretical RGB
width, obtained from appropriate theoretical models, as a function of
metallicity and age; this allows us to properly compare, in
Section~\ref{sec:compar}, the RGB width properties of the Galactic and
extragalactic GCs in our database; in Section~\ref{sec:age} we analyze in
detail the relation between the RGB width and the age of the clusters. Finally,
Section~\ref{sec:summary} provides a summary of the main results.

\section{Observations and data reduction}\label{sec:obs}
The cluster database analyzed in this work is composed of Galactic and
extragalactic GCs. The Galactic sample includes the 57 GCs present in
\citet{milone17} plus the GC IC\,4499 \citep{milone18b}, and has been
complemented with the Galactic GCs \ngc2419 \citep{zennaro19} and Ruprecht\,106
\citep[Rup\,106;][]{dotter18}. The extragalactic database includes: six Small
Magellanic Cloud (SMC) clusters, namely \ngc121, \ngc339, \ngc416, Lindsay\,1,
Lindsay\,38, and Lindsay\,113; the Large Magellanic Cloud (LMC) \ngc1978; the
cluster Terzan\,7 associated to the Sagittarius Dwarf spheroidal
\citep{sbordone05}. Details about photometric reduction, differential reddening
correction and cluster membership selection can be found in \citet{milone17}
and \citet{milone18b} for the 58 Galactic GCs, in \citet{zennaro19} for
\ngc2419 and \citet{dotter18} for Rup\,106, and in \citet{lagioia19} for the
SMC clusters \ngc121, \ngc339, \ngc416, and Lindsay\,1. Archival images of the
remaining GCs, namely Lindsay\,38, Lindsay\,113, \ngc1978, collected through
the Wide Field Channel of the Advanced Camera for Surveys (WFC/ACS)
and the Ultraviolet and Visual Channel of the Wide Field Camera 3 (WFC3/UVIS)
onboard \hst, and archival images of Terzan\,7 collected through Focal
Reducer/low dispersion Spectrograph 2 (FORS2) at the Very Large Telescope
(VLT), have been specifically processed for this work, according to the
procedures described in the following section. Observation details about these
four clusters are reported in Table~\ref{tab:obs}.

\begin{deluxetable*}{*{6}{c}}
\tabletypesize{\small}
\tablewidth{0pt}
\tablecaption{Observations of Lindsay\,38, Lindsay\,113, \ngc1978, and Terzan\,7.\label{tab:obs}}
\tablehead{
\colhead{Cluster} & \colhead{Date} & \colhead{Camera} & \colhead{Filter} & \colhead{N$\times$exposure time (s)} & \colhead{Program ID}
}
\startdata
Lindsay\,38  & Jun 03 2004 -- 18 Aug 2005 & WFC/ACS   & F555W               & $2\times20 + 480 + 4\times485$                                & 9891,10396   \\            
	     & Jun 03 2004 -- 18 Aug 2005 &    ''     & F814W               & $2\times10 + 290 + 4\times463$                                &    ''        \\              
	     & Oct 21 2017                & WFC3/UVIS & F336W               & $268 + 2\times710$                                            & 15062        \\
	     & Oct 21 2017                &    ''     & F343N               & $515 + 2\times1057$                                           &    ''        \\
	     & Oct 21 2017                &    ''     & F438W               & $123 + 2\times538$                                            &    ''        \\
Lindsay\,113 & Jun 03 2004                & WFC/ACS   & F555W               & $480$                                                         & 9891         \\
             & Jun 03 2004                &    ''     & F814W               & $290$                                                         &    ''        \\
	     & Sep 05 2018                & WFC3/UVIS & F336W               & $274 + 2\times720$                                            & 15062        \\
             & Sep 05 2018                &    ''     & F343N               & $530 + 2\times1065$                                           &    ''        \\
             & Sep 05 2018                &    ''     & F438W               & $128 + 2\times545$                                            &    ''        \\
 \ngc1978    & Oct 07 2003                & WFC/ACS   & F555W               & $300$                                                         & 9891         \\
             & Oct 07 2003                &    ''     & F814W               & $200$                                                         &    ''        \\
             & Aug 15 2011 -- Sep 25 2016 & WFC3/UVIS & F336W               & $380 + 460 + 660 + 740$                                       & 12257, 14069 \\
             & Aug 15 2011                &    ''     & F555W               & $60 + 300 + 680$                                              & 12257        \\
             & Sep 25 2016                &    ''     & F343N               & $425 + 450 + 500 + 2\times800 + 1000$                         & 14069        \\
             & Sep 25 2016                &    ''     & F438W               & $75 + 120 + 420 + 460 + 650 + 750$                            &    ''        \\
Terzan\,7    & Apr 14 2004                & FORS2@VLT & $\mathrm{B_{BESS}}$ & $1\times30 + 2\times120$                                      & 073.D-0273 \\
             & Mar 18 2006                &    ''     & $\mathrm{I_{BESS}}$ & $1\times0.3 + 1\times3 + 1\times30$                           & 077.D-0775 \\
             & Mar 18 -- 22 2011          &    ''     & $\mathrm{U_{HIGH}}$ & $6\times5 + 1\times21 + 6\times 60 + 21\times94 + 1\times133$ & 087.D-0290 \\
%
%
\enddata
\end{deluxetable*}
%
%

\subsection{Lindsay\,38, Lindsay\,113 and \ngc1978}

The images of the clusters Lindsay\,38, Lindsay\,113 and \ngc1978, available at
the \hst\ MAST archive\footnote{\url{http://archive.stsci.edu/hst/}}, have been
reduced following the procedure described in \citet{anderson08} and used by
\citet{lagioia19} for the analysis of the aforementioned four SMC GCs. Shortly,
the data reduction was performed on the \textit{flt} images after applying the
correction for poor Charge Transfer Efficiency \cite{anderson10a}. For each
scientific frame, an array of Point Spread Functions (PSFs) has been computed,
starting from  library empirical PSFs and adding to each PSF a spatial
variation correction computed from isolated, unsaturated bright stars. The
\hst\ image reduction software \textsc{img2xym} \citep{anderson06} was used to
detect bright stars and accurately determine their position and flux. This
program takes also into account the amount of star's flux bled into adjacent
pixels \citep{anderson08,gilliland04,gilliland10}. The position and flux of
faint stars was obtained with a different program (\textsc{KS2}; J. Anderson et
al., in preparation), which combines the information of position and flux of
each star in all the frames in which a star's image in present. Details about
this algorithm can be found in \citet{sabbi16} and \cite{bellini17}. The
instrumental magnitudes have been calibrated to the VEGAMAG system according to
\citet{bedin05}. The encircled energy distribution corrections and Zero Points
of the UVIS detector of the WFC3 camera have been taken from the STScI
website\footnote{\url{http://www.stsci.edu/hst/wfc3/analysis/uvis_zpts/},/
\url{http://www.stsci.edu/hst/acs/analysis/zeropoints}}. Stellar positions have
been corrected for geometric distortion by using the solution provided by
\citet{bellini11}. As a final step, for our analysis we selected the
best-measured stars by employing the photometric quality indexes provided by
the software \citep[see][]{milone09a}. 

Finally, we found that, by applying the method of \citet{milone12c}, the effect
of differential reddening on the photometry of Lindsay\,38 and \ngc1978 is
negligible, being below the typical photometric errors. No correction for
differential reddening has been applied, therefore, to these two clusters.  On
the other hand, with the same method, we estimated the reddening variation in
the  field of view (FoV) of Lindsay\,113 and found $\Delta \mathrm{E(B-V)}$
ranging within $\pm 0.020$\,mag, with 68.27\% of stars having $\Delta
\mathrm{E(B-V)} < 0.005$\,mag.  We applied this correction to the magnitudes of
Lindsay\,113 using the following relations for the total-to-differential
absorption: $\mathrm{A_{F336W} = 5.100\,E(B-V)}$, $\mathrm{A_{F438W} =
4.182\,E(B-V)}$, and $\mathrm{A_{F814W} = 1.842\,E(B-V)}$ \citet[][private
communication]{dotter16}.

\subsection{Terzan 7}
For this cluster we used $\mathrm{U_{HIGH}, B_{BESS}}$ and $\mathrm{I_{BESS}}$
band images available at the ESO
archive~\footnote{\url{archive.eso.org/cms.html}}.  The observation dataset of
Terzan 7 includes 41 exposures, of which 35 in the $\mathrm{U_{HIGH}}$ band,
three in the $\mathrm{B_{BESS}}$ band, and three in the $\mathrm {I_{BESS}}$
band. All the exposures were collected with
the High-Resolution collimator, resulting in a reduced total FoV of $\sim
4.25\times 4.25$ arcmin and a pixel scale of $\sim 0.125$ arcsec/pixel.
Additional details concerning the specific observations are provided in
Table~\ref{tab:obs}. 

Every FORS2 frame is composed of two images, one for each of the two chips
making up the FORS2 detector. The exposures have been collected following a
dithering pattern which includes the cluster core radius
\citep[0.77 arcmin;][2010 update]{harris96a} in the upper part of the detector
or `chip 1'. Pre-reduction of science data, consisting in bias subtraction and
subsequent flat-field correction, was performed with the pipeline
\textsc{EsoReflex} \citep{freudling13}. The reduction workflow of this program
automatically selects, for each science frame, the correct calibration frames
listed in the association (\texttt{.xml}) files retrieved from the archive
together with science FITS files. Finally, every pre-reduced frame has been
split into two portions, corresponding to the `chip 1' and `chip 2' (lower
chip) images. Each image has been then analyzed separately.

Point-spread function (PSF) photometry has been performed with \textsc{DAOPHOT}
and \textsc{ALLSTAR} \citep{stetson87,stetson94}. For each image, the
computation of the best PSF model has been carried out by selecting 40 to 50
high S/N stars, subtracting the contribution of the neighbor stars to the their
luminosity profile and computing the resulting PSF model. The procedure was
repeated by increasing, at every step, the degrees of freedom of the spatial
variation of the PSF across the frame, starting from a pure analytical model
and ending with a quadratically varying PSF. By matching all the single star
lists with appropriate geometric transformation we obtained a master list
containing all the stars measured at least once in the various images. The
geometric transformation solution has been found with \textsc{DAOMATCH} and
\textsc{DAOMASTER} \citep{stetson90}. Then, by running \textsc{ALLFRAME}
\citep{stetson94}, every star in the master list that was present, according to
the geometric transformation, in each single frame, has been measured again and,
for each image a new list of improved position and magnitudes obtained. 

For each filter, the new star lists have then been matched, obtaining three
catalogs of instrumental position and magnitude in U, B and I, respectively.
Finally the position of the B-band and I-band catalogs have been transformed to
the reference system of the U-band one.

Unfortunately no standard-star fields in U band have been observed for both the
U-band observing runs, namely on 18 and 22 March 2011. Furthermore, to our
knowledge no catalogs including secondary photometric standard stars in this
passband are available for the FoV covered by the observation dataset. Since,
as a consequence, it was not possible to calibrate the images in U band, we
performed the following analysis using instrumental magnitudes. We notice,
however, that our analysis is based on the measurement of the cluster RGB width
which is a relative quantity and, as such, it is negligibly affected by the
use of calibrated magnitudes. 

For our analysis, we selected the best measured stars by using the photometry
quality parameters provided by the software, namely $sharp$ and $chi$
\citep{stetson87}. In particular we rejected stars with $|sharp| \geq 0.15$ and
$chi > 0.9$. 

Finally, since the effect of differential reddening on the photometry
of this cluster is negligible, no correction has been applied.

\subsection{Cluster membership}
For the Galactic GCs in our database, the cluster membership of stars has been
assessed on the basis of their proper motions, as described in
\citet{milone17,milone18b}, to which we refer the interested reader. For all
the other clusters, we adopted the same procedure used in the analysis of
\citet{lagioia19}, based on the star's position in the observed FoV: we
empirically determined the center of each cluster by eye and then selected all
the stars within 0.67\,arcmin from the cluster center.  Only in the case of
Terzan\,7, we flagged as cluster members all the stars within the cluster
half-light radius \citep[0.77 arcmin;][2010 update]{harris96a}, which are all
located in the `chip 1'. In this case, the radial selection, about a factor of
$\sim 1.2$ times larger than the previous one, compensates for the lower number
of detections in the central cluster region, due to the fact that crowding in
the cluster core is critical for ground-based observations. We notice that the
adopted radial selection includes the central region of every cluster, thus
resulting in a marginal field star contamination. 

In the following we refer our analysis only to the cluster member stars. For
each GC, we tagged as RGB only those stars lying along the observed RGB
sequence in all the different UV and optical color combinations, in order to
exclude possible contamination from Asymptotic Giant Branch (AGB) and
Horizontal Branch (HB) stars.

\begin{figure*}
\centering
\includegraphics[angle=270,width=\textwidth]{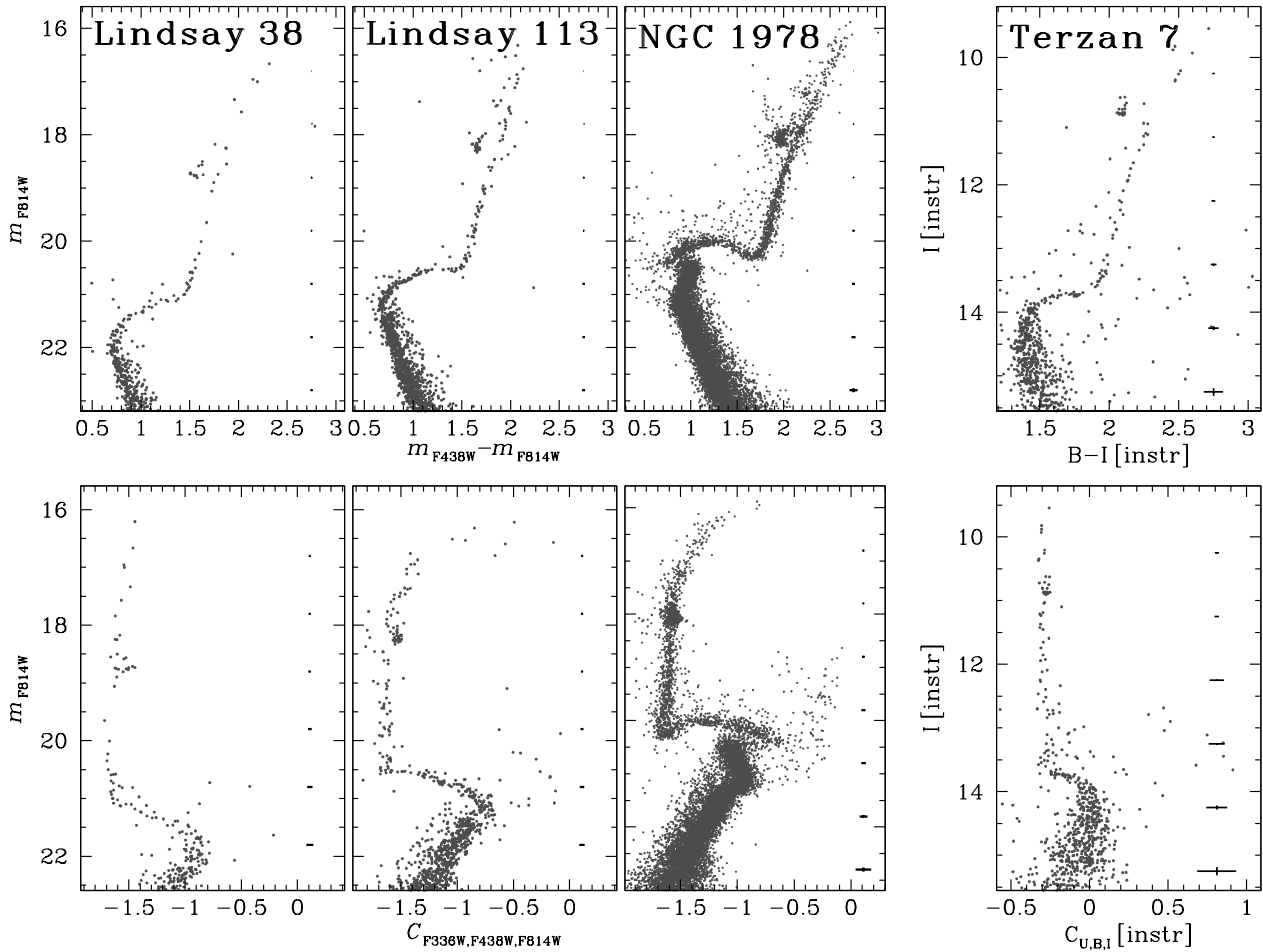} 
\caption{$m_{F814W}$ vs. $(m_{F336W}-m_{F438W})$ CMD (top panels) and
	$m_{F814W}$ vs. $C_{F336W,F438W,F814W}$ pseudo CMD (bottom panels) of
	the cluster members of the SMC globulars Lindsay\,38, Lindsay\,113, and
	the LMC globular \ngc1978. The instrumental $I$ vs ($B-I$) CMD (top) and
	$I$ vs $C_{U,B,I}$ (bottom) pseudo-CMD of the cluster members of the
	Sagittarius Dwarf cluster Terzan\,7 are plot in the right panels.
	In each panel, the black error bars indicate the typical photometric
	errors at different luminosities.\label{fig:cmds}} 
\end{figure*}

The final CMDs of Lindsay\,38, Lindsay\,113, \ngc1978 and Terzan\,7 have been
plotted in  Figure~\ref{fig:cmds} which displays the $m_{F814W}$ vs.
$(m_{F438W}-m_{F814W})$ CMDs and $m_{F814W}$ vs.  $C_{F336W,F438W,F814W}$
pseudo-CMDs of the cluster members, respectively in the top and bottom panels.
In the case of Terzan\,7 the instrumental $I$ vs. $(B-I)$ CMD and $I$ vs.
$C_{U,B,I}$ pseudo-CMD are displayed. A glance at the
CMDs reveals that the total number of RGB stars considerably changes from one
cluster to another, with Lindsay\,38 and \ngc1978 being, respectively, the
least and most populated cluster. The error bars plotted on the right side of
each panel indicate the typical photometric errors along the entire magnitude
extension of the CMD.

\subsection{Global cluster parameters}
This work includes the study of the monotonic relationship between the RGB
width of the Galactic GCs and a set of global GC parameters (see
Sect.~\ref{sec:corr}) composed by 45 observational, morphological and dynamical
quantities. In particular, we obtained: from \citet[][2010 update]{harris96a}:
metallicity (\feh), reddening ($\mathrm{E(B-V)}$), absolute visual magnitude
($\mathrm{M_V}$), central surface brightness (SB$_0$), central luminosity
density ($\rho_0$), projected ellipticity of isophotes (epsilon), and
concentration (c); from \citet[][and references therein]{baumgardt18b} total
cluster mass (Mass), mass-to-light ratio in V band (M/L), cluster core radius
(r$_c$), projected half-light radius (r$_{hl}$), half-mass radius (r$_{hm}$),
tidal radius (r$_t$), core density ($\rho_c$) half-mass radius density
($\rho_{hm}$), half-mass relaxation time ($T_{RH}$), slope of mass function
(MF slope), mass fraction of remnants ($\mathrm{F_{remn}}$),
central velocity dispersion ($\sigma_0$), central escape velocity ($v_{esc}$),
mass segregation parameter for stars within 0.5 and 0.8 $M\odot$ inside the
core ($\eta_{c}$) and inside the half-mass radius ($\eta_{hm}$); from
\citet[][and references therein]{baumgardt19}: mean heliocentric radial
velocity ($\mathrm{\langle RV \rangle}$), X, Y, Z component of the cluster
position and the corresponding velocity component U, V, W, distance from the
center of the Galaxy ($\mathrm{R_{GC}}$), perigalacticon
($\mathrm{R_{perig}}$), apogalacticon ($\mathrm{R_{apog}}$); from
\citet{marin-franch09}: relative age
($\mathrm{age_{\,\text{\citetalias{marin-franch09}}}}$) based on \citet{dotter07}
models in the \citet{carretta97} scale; absolute ages both from
\citet{dotter10,dotter11} and \citep{milone14}
($\mathrm{age_{\,\text{\citetalias{dotter10}}}}$) and \citet{vandenberg13}
($\mathrm{age_{\,\text{\citetalias{vandenberg13}}}}$); from \citet{milone18b}:
mean and maximum internal helium variation ($\mathrm{\langle \delta Y
\rangle}$, $\mathrm{\delta Y_{max}}$); from \citet{milone17}: width of the RGB
measured in the pseudo-color $C_{F275W,F336W,F438W}$
($W_{C}$(\citetalias{milone17})), ratio of 1G stars to total number of stars
($\mathrm{N_{1G}/N_{tot}}$); from the 2003 edition of the Harris catalog of GCs
\citep{harris96a}: specific frequency of RR Lyr\ae\ (S(RR\,Lyr)); from
\citet{milone12c}: fraction of binaries in the cluster core
($\mathrm{F_{bin(c)}}$), between the cluster core and the
half-mass radius ($\mathrm{F_{bin(hm)}}$), and beyond the
half-mass radius ($\mathrm{F_{bin(o-hm)}}$); from \citet{mackey05}: horizontal
branch ratio (HBR); from \citet{milone14}: extension of
($m_{F606W}-m_{F814W}$) color of the HB (L2).

Masses and structural parameters by \citet{baumgardt18b} have been obtained
from N-body simulations based on velocity dispersion and surface density
profiles deriving from $\sim 35,000$ archival radial velocities from ESO/VLT
and Keck. The orbital parameters by \citet{baumgardt19} have been obtained by
cross-matching the aforementioned radial velocities and GAIA Data Release 2
proper motions \citep{gaiadr218}.

\section{Measurement of the intrinsic RGB width \label{sec:rgbw}} 
In this section we describe the procedure to measure the RGB width in the
pseudo-color $C_{F336W,F438W,F814W}$, for each GC in our database, using the
SMC cluster \ngc416 as a template to illustrate the methodology. 

\begin{figure*}
\centering
\includegraphics[width=.9\textwidth]{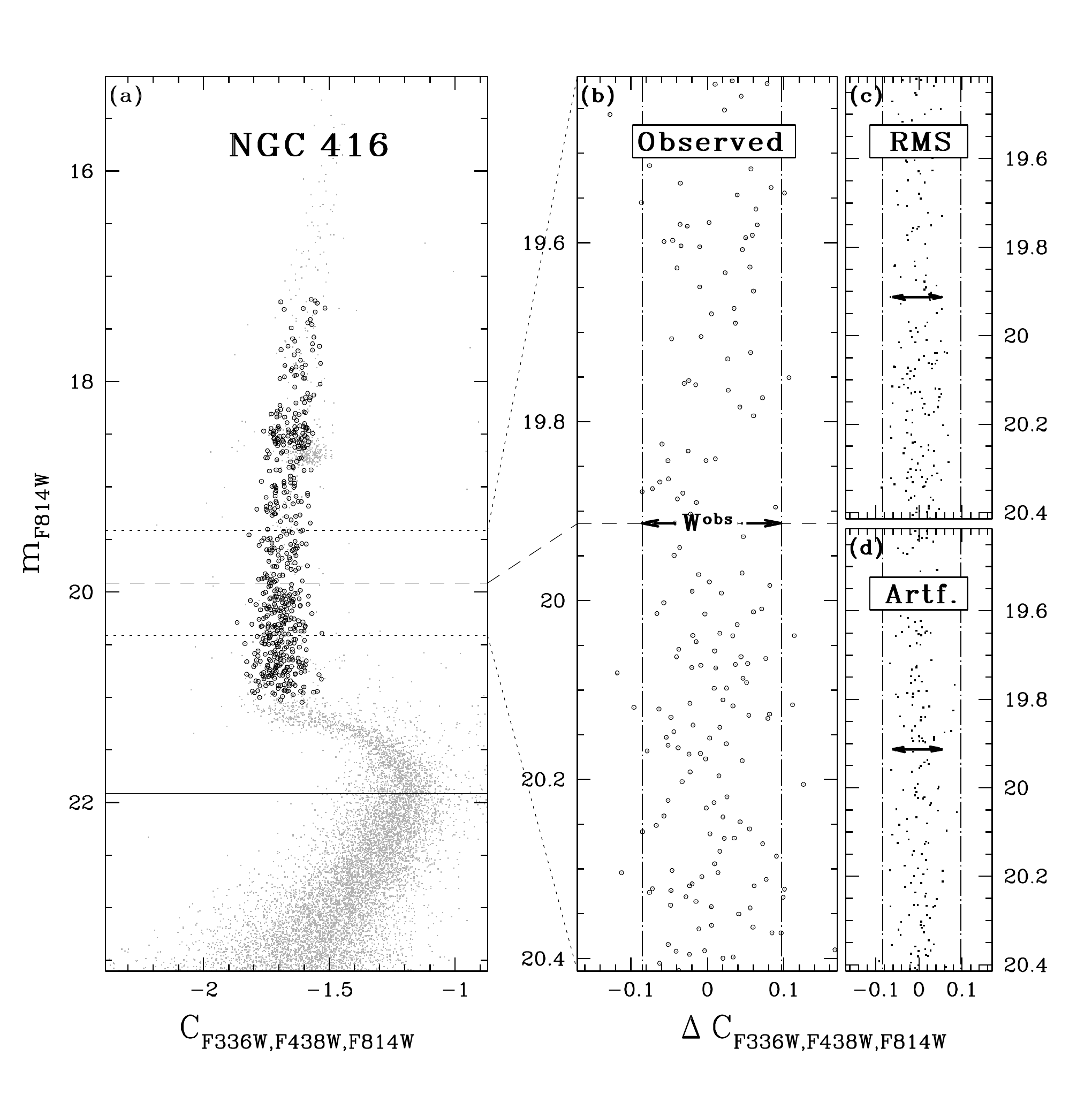} 
\caption{Procedure to measure the intrinsic RGB width of the cluster \ngc416.
	\textit{(a)}: $m_{F814W}$ vs. $C_{F336W,F438W,F814W}$ pseudo
	CMD, where RGB stars are represented as black dots. The MSTO
	and the reference point at $m^{MSTO}_{F814W}-2$ are indicated,
	respectively, by the solid and dashed horizontal line, while the dotted
	lines delimit 1\,mag interval around the reference point. \textit{(b)}:
	verticalized CMD for the RGB stars in the 1\,mag interval: the $4^{th}$
	and $96^{th}$ percentile of their color distribution are indicated as
	dot-dashed vertical lines.  The distance between these two lines
	represents the observed RGB width $W^{obs}_{C F336W,F438W,F814W}$ and,
	for simplicity purpose, it has been indicated as $W^{obs}$ in the
	figure. \textit{(c)}: artificial distribution of standard color errors,
	used to estimate the contribution of the photometric error to the
	observed RGB width. The RGB width is indicated by the double
	arrow.  \textit{(d)}: same as in panel (c) but using artificial
	stars.\label{fig:w}}
\end{figure*}

Panel (a) of Figure~\ref{fig:w} displays the $m_{F814W}$ vs
$C_{F336W,F438W,F814W}$ CMD of \ngc416, with the cluster stars represented as
gray dots. The first step consisted in the measurement of the main sequence
turn-off (MSTO) luminosity, $m^{MSTO}_{F814W}$, represented as a solid
black line in the plot. The quantity $m^{MSTO}_{F814W}$ has been determined by
using the naive estimator of \citet{silverman86}, which consists in: i) the
subdivision of the analyzed magnitude range in a given number of magnitude bins
and the calculation of the median color and magnitude of the stars included in
each bin; ii) the implementation of the previous algorithm for different bin
series, obtained by shifting the starting point defining the first bin, of a
quantity equal to a fraction of the predefined bin width. The magnitude and
color of the resulting median points has been smoothed by boxcar averaging
three adjacent points, and the bluest color of the linear function
interpolating the resulting smooth function has been taken as the
$m^{MSTO}_{F814W}$ estimate. The following step was the definition of a
luminosity interval of 1\,mag centered around the reference luminosity defined
as $m^{MSTO}_{F814W}-2$ \citep[see][]{milone17} and indicated as a dashed line.
The two dotted lines in the plot delimit the selected magnitude interval. At
this point, we computed the RGB fiducial line of the RGB stars, by using the
method described above, and measured the pseudo-color difference between each
RGB stars and the fiducial line at the same F814W magnitude ($\Delta
C_{F336W,F438W,F814W}$). Panel (b) in Fig.~\ref{fig:w} shows $m_{F814W}$
against $\Delta C_{F336W,F438W,F814W}$.  The observed RGB width, $W^{obs}_{C
F336W,F438W,F814W}$, corresponds to the difference between the $96^{th}$ and
$4^{th}$ percentile (vertical lines in the plot) of the $\Delta
C_{F336W,F438W,F814W}$ distribution, and is equal, in the case of \ngc416, to
0.180\,mag. 

The error on the estimate of the observed width was obtained by carrying out
10,000 bootstrapping tests on random sampling with replacement of the
observed RGB stars in the selected magnitude interval. Each test
consists in the generation of an artificial sample containing 1,000 sequential
copies of the observed stellar colors and the subsequent random extraction of a
subsample composed of a number of colors equal to the observed one. For each
extraction we then computed the corresponding RGB width. The $68.27^{th}$
percentile of the resulting distribution of the 10,000
bootstrapping measurements was taken as the standard error of $W^{obs}_{C
F336W,F438W,F814W}$, equal to 0.013\,mag for \ngc416.

$W^{obs}_{C F336W,F438W,F814W}$, however, does not correspond to the
\textit{intrinsic} width of the cluster RGB, because the photometric errors
concur to the observed broadening of the RGB. In order to quantify
such spurious contribution, we took advantage of the magnitude r.m.s. of the
stars in the selected magnitude interval. For each cluster, we
simulated a series of 10,000 artificial CMDs, with a random color spread based
on the observed color errors. To do this we assigned to each observed
star in the selected magnitude interval, an artificial error, in each of the
three bands F336W, F438W and F814W. The artificial error was obtained by
extracting a random value from a simulated Gaussian error distribution with a
standard deviation equal to the observed star's magnitude standard error,
derived from the reduction software. We then measured the resulting artificial
RGB width and subtracted it in quadrature from the observed RGB width,
obtaining an estimate of the intrinsic width. We repeated this procedure 10,000
times and took the mean value resulting from the whole set of
simulations as the intrinsic width of the cluster. We obtained for
\ngc416 $W_{C F336W,F438W,F814W} = 0.146$\,mag. As an example, in panel (c), we
plotted the $m_{F814W}$ vs.\,$\Delta C_{F336W,F438W,F814W}$ diagram, for one
simulation, and indicated the corresponding RGB width with the double
arrow.  For a quick comparison, we plotted again the observed $4^{th}$ and
$96^{th}$ percentile vertical lines in this panel.

Artificial stars (ASs) provide an independent estimate of the contribution of
the photometric error. To obtain this estimate, we applied the same procedure
described above but using a randomly extracted sub-sample of ASs composed of
the same number of observed stars and derived the corresponding $\Delta
C_{F336W,F438W,F814W}$ value, as shown in panel (d) of Figure~\ref{fig:w}.
The latter value represents the difference between the input and output
pseudo-color, obtained from the ASs test \citep[see][and references
therein]{lagioia19}. We obtained $W_{C F336W,F438W,F814W} = 0.147$\,mag, the
same value within 0.001\,mag as that obtained by using the star's r.m.s. 

We applied the above procedure to compute the intrinsic RGB width of all the 68
clusters analyzed in this work. Since AS catalogs are, at the present time,
only available for the MC GCs in our database, for these clusters only we used
the ASs for the calculation of the photometric error contribution.  Moreover,
since Lindsay\,113 shows a sparsely populated RGB, for this GC we repeated the
procedure for the estimate of the intrinsic RGB width by excluding first the
bluest and then also the reddest RGB stars in the selected magnitude interval. The
average of the two RGB width estimates was taken as the final RGB width, while
the half interval between them as the associated error.  The intrinsic RGB
values of all the analyzed clusters have been reported in Table~\ref{tab:ws}.
We see that the $W_{C F336W,F438W,F814W}$ estimates range from a minimum value
of 0.034 mag (Lindsay\,38) to a maximum value of 0.332\,mag (\ngc6441) and that
the majority of GCs are distributed in the interval 0.1 -- 0.2.

\begin{deluxetable*}{lr@{$\,\pm\,$}lDlr@{$\,\pm\,$}lD}
\tablewidth{10cm}
	\tablecaption{Intrinsic and `normalized' RGB width of the analyzed Globular Clusters.\label{tab:ws}}
\tablehead{
   \colhead{ID} & \multicolumn2c{$W_{C F336W,F438W,F814W}$} & \multicolumn2c{$\Delta W_{C F336W,F438W,F814W}$} \
 & \colhead{ID} & \multicolumn2c{$W_{C F336W,F438W,F814W}$} & \multicolumn2c{$\Delta W_{C F336W,F438W,F814W}$} \\
   \colhead{} & \multicolumn2c{(mag)} & \multicolumn2c{(mag)} & \colhead{} & \multicolumn2c{(mag)} & \multicolumn2c{(mag)} 
}
\decimals
\startdata
IC\,4499      & 0.118 & 0.018 & -0.012 &  \ngc6205      & 0.154 & 0.005 &  0.024 \\
LINDSAY\,1    & 0.135 & 0.009 & -0.023 &  \ngc6218      & 0.156 & 0.008 &  0.014 \\
LINDSAY\,38   & 0.034 & 0.014 & -0.098 &  \ngc6254      & 0.151 & 0.008 &  0.023 \\
LINDSAY\,113  & 0.108 & 0.037 & -0.031 &  \ngc6304      & 0.222 & 0.010 &  0.016 \\
\ngc104       & 0.205 & 0.006 &  0.018 &  \ngc6341      & 0.115 & 0.006 &  0.040 \\
\ngc121       & 0.157 & 0.008 &  0.011 &  \ngc6352      & 0.198 & 0.012 &  0.005 \\
\ngc288       & 0.148 & 0.028 &  0.003 &  \ngc6362      & 0.174 & 0.011 &  0.006 \\
\ngc339       & 0.075 & 0.004 & -0.084 &  \ngc6366      & 0.173 & 0.022 & -0.023 \\
\ngc362       & 0.155 & 0.005 &  0.006 &  \ngc6388      & 0.303 & 0.007 &  0.104 \\
\ngc416       & 0.151 & 0.013 & -0.019 &  \ngc6397      & 0.103 & 0.031 &  0.007 \\
\ngc1261      & 0.145 & 0.006 &  0.002 &  \ngc6441      & 0.332 & 0.011 &  0.126 \\
\ngc1851      & 0.173 & 0.010 &  0.027 &  \ngc6496      & 0.173 & 0.019 & -0.033 \\
\ngc1978      & 0.129 & 0.009 & -0.084 &  \ngc6535      & 0.085 & 0.031 & -0.027 \\
\ngc2298      & 0.116 & 0.018 &  0.013 &  \ngc6541      & 0.142 & 0.006 &  0.031 \\
\ngc2419      & 0.205 & 0.009 &  0.118 &  \ngc6584      & 0.153 & 0.005 &  0.021 \\
\ngc2808      & 0.199 & 0.004 &  0.041 &  \ngc6624      & 0.234 & 0.008 &  0.027 \\
\ngc3201      & 0.153 & 0.014 &  0.027 &  \ngc6637      & 0.197 & 0.007 &  0.004 \\
\ngc4590      & 0.104 & 0.017 &  0.023 &  \ngc6652      & 0.174 & 0.008 & -0.007 \\
\ngc4833      & 0.126 & 0.007 &  0.018 &  \ngc6656      & 0.148 & 0.012 &  0.064 \\
\ngc5024      & 0.117 & 0.005 &  0.027 &  \ngc6681      & 0.147 & 0.008 &  0.023 \\
\ngc5053      & 0.087 & 0.020 &  0.009 &  \ngc6715      & 0.189 & 0.013 &  0.085 \\
\ngc5139      & 0.184 & 0.006 &  0.131 &  \ngc6717      & 0.154 & 0.022 &  0.005 \\
\ngc5272      & 0.131 & 0.005 & -0.001 &  \ngc6723      & 0.210 & 0.007 &  0.049 \\
\ngc5286      & 0.207 & 0.012 &  0.102 &  \ngc6752      & 0.162 & 0.011 &  0.032 \\
\ngc5466      & 0.078 & 0.006 & -0.021 &  \ngc6779      & 0.137 & 0.009 &  0.038 \\
\ngc5897      & 0.117 & 0.031 &  0.013 &  \ngc6809      & 0.106 & 0.012 &  0.005 \\
\ngc5904      & 0.192 & 0.015 &  0.045 &  \ngc6838      & 0.167 & 0.011 & -0.016 \\
\ngc5927      & 0.245 & 0.013 &  0.042 &  \ngc6934      & 0.185 & 0.011 &  0.045 \\
\ngc5986      & 0.180 & 0.009 &  0.054 &  \ngc6981      & 0.131 & 0.005 & -0.007 \\
\ngc6093      & 0.147 & 0.008 &  0.032 &  \ngc7078      & 0.109 & 0.004 &  0.038 \\
\ngc6101      & 0.092 & 0.016 & -0.007 &  \ngc7089      & 0.152 & 0.007 &  0.032 \\
\ngc6121      & 0.123 & 0.006 & -0.033 &  \ngc7099      & 0.098 & 0.015 &  0.020 \\
\ngc6144      & 0.114 & 0.010 &  0.000 &  RUPRECHT\,106 & 0.061 & 0.009 & -0.059 \\
\ngc6171      & 0.175 & 0.013 &  0.009 &  TERZAN\,7     & 0.043 & 0.009 & -0.172 \\
\enddata
\end{deluxetable*}

\section{Bivariate analysis of the RGB width} \label{sec:corr}
We performed an extensive statistical analysis to test the monotonic
relationship between the cluster intrinsic RGB width and several GC parameters,
using as indicator the Spearman's rank correlation coefficient, R$_S$. 
The following run of statistical tests has been done by taking into account a
sub-sample of clusters in our database, that includes the 58 Galactic GCs used
by \citet{milone18b}. 

\begin{figure*}
\centering
\includegraphics[width=\textwidth]{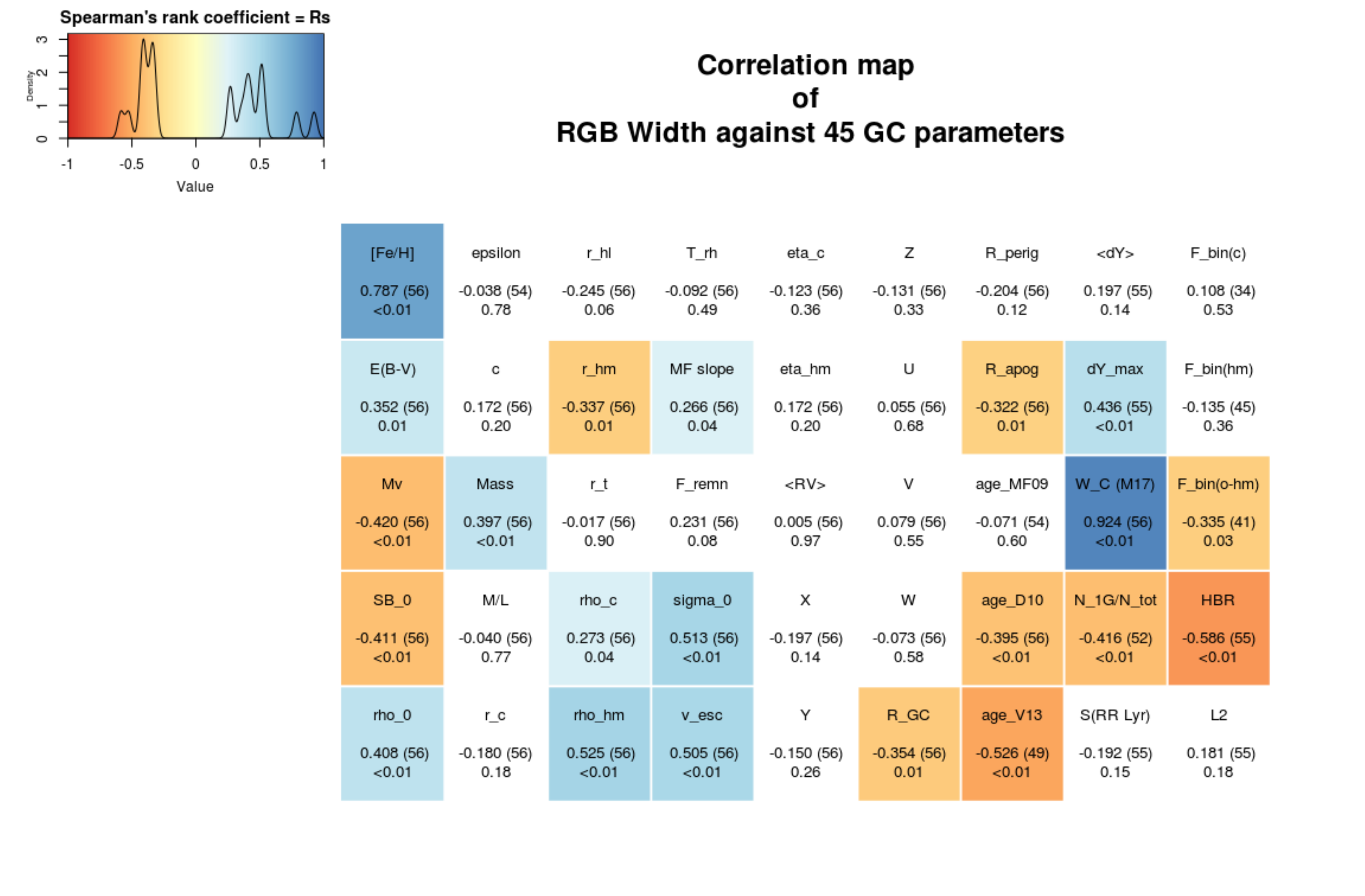} 
\caption{Correlation map of the Spearman's rank correlation test of the RGB
	width ($W_{C F336W,F48W,F814W}$) of Galactic GCs against a set of 45 GC
	parameters. Each cell represents the result of the test against a
	single parameter and is labeled with the name of the corresponding
	parameter (top row), the value of the Spearman's rank correlation coefficient
	R$_S$, followed by the system degrees of freedom in parentheses (middle
	row), and the significance of the correlation measurement indicated as
	the p-value of the coefficient (bottom row). Each cell is color-coded
	on a scale, shown in the top-left corner, which maps R$_S$ into a color
	varying from red (R$_S = -1$) to blue (R$_S = +1$). No color has been
	assigned to cells of non significant (p-value $>0.05$) correlation
	coefficients. The kernel distribution of the significant measurements
	has been overplot to the color-key scale. \label{fig:corrW_map}} 
\end{figure*}

The results are shown as a correlation map in Figure~\ref{fig:corrW_map}. Each
cell in the figure represents the test scores for a single parameter, and is
labeled with three rows. The top row reports the parameter against which the
Spearman's correlation test has been performed, the middle row reports the
result of the test, namely the value of R$_S$, followed by a number in
parentheses equal to the degrees of freedom of the system, calculated as the
pairwise cases used for the test minus two. The bottom row reports the
significance of the R$_S$ measurement, indicated as the p-value or the
probability to find a R$_S$ value equal or larger than the actual one. Strong
evidence against no correlation is usually given by p-values $\leq 0.05$, while
p-values smaller than 0.01 mark highly significant correlations and have been
reported as an upper limit value, namely `$< 0.01$'. The coefficients R$_S$
have been mapped into colors, assigned to the relative cells, on a scale
reported in the top-left corner of the figure, where the red corresponds to a
perfect negative correlation (R$_S = -1$) while the blue to a perfect positive
correlation (R$_S = +1$). Moderate to weak correlations ($-0.6 \lesssim$ R$_S
\lesssim +0.6$) are represented by yellowish colors. Finally no color has been
assigned to cells whose test scores are not significant, namely with p-value $>
0.05$, because no conclusions can be drawn from the measured correlations. 

The map shows that 21 out of 45 analyzed parameters have a significant
correlation with the RGB width. The kernel density distribution of the
significant measurements, overplot to the color-key scale, shows that almost
all the R$_S$ lie in the interval $\sim -0.5, \sim +0.5$, except for
$W_{C}$(\citetalias{milone17}) (R$_S[56] = 0.924$, p $ < 0.01$) and for \feh\
(R$_S[56] = 0.787$, p $ < 0.01$). 

The first result is of remarkable importance since the almost perfect
monotonic correlation between the indexes $W_{C F336W,F438W,F814W}$ and $W_{C
F275W,F48W,F814W}$, means that both are equally sensitive to variations of
light elements (C and N) in MPs.  This finding also demonstrate of the
advantage of the first index over the second in MPs studies, given its high
effectiveness/cost ratio.

The second result, similar to that obtained in \citet{milone17}, indicates that
metallicity is the most important physical parameter impacting on the extension
of the RGB width. It is therefore necessary to take into account the effect of
the variance of metallicity on the rest of parameters to verify if their
correlation with the RGB width still holds.

\begin{figure*}
\centering
\includegraphics[width=\textwidth]{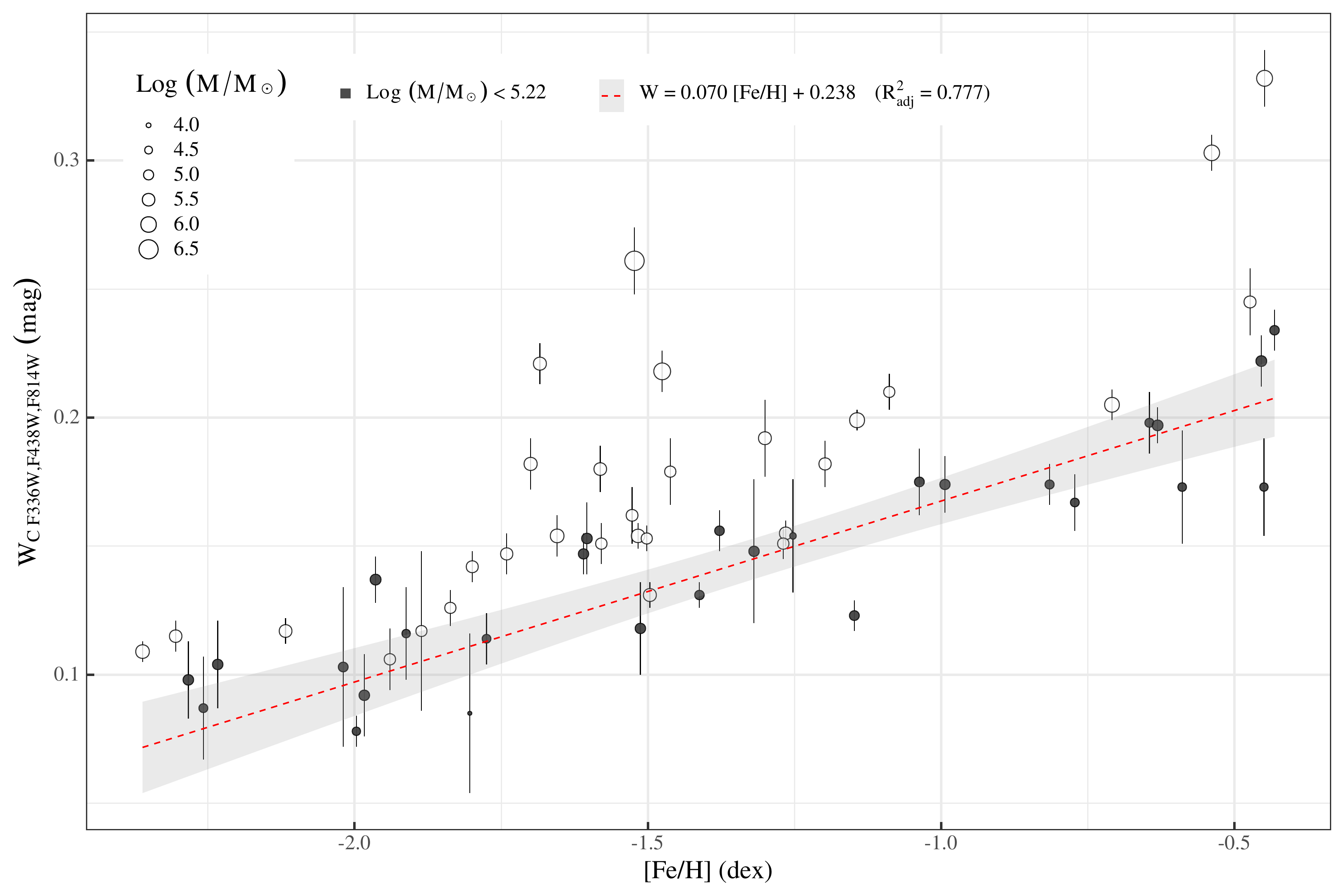} 
\caption{RGB width $W_{C F336W,F438W,F814W}$ vs. \feh\ for the Galactic GCs in
	our database. The dimension of each point is proportional to the mass
	of the corresponding cluster. The linear relation fitting the clusters
	with Log\,(M/M$_{\odot}) < 5.22$, marked by the black-filled
	points, and used to remove the effect of metallicity from the RGB width
	of all the clusters in our database, is represented by the red
	dashed line. The gray-shaded area marks the 95\% confidence interval of
	the true regression line.  The linear fit relation and the
	adjusted determination coefficient R$_{adj}^2$ are also reported in
	the legend.\label{fig:lfit}} 
\end{figure*}

Figure~\ref{fig:lfit} shows the scatter plot of $W_{C F336W,F438W,F814W}$ vs.
\feh. The vertical error bar of each point represents the error of the RGB
width measurement (see table~\ref{tab:ws}). For the sake of clarity, in order
to avoid overlap with points and error bars, we added to the metallicity values
of the plotted points a small ($\leq 0.02$ dex) random jitter. As it follows
from the previous result, an evident monotonic progress of the points is
visible, with the RGB width increasing with metallicity. The overall trend also
suggests a moderate degree of linearity between the two represented quantities,
with a significant scatter in the metallicity interval $-1.7 \lesssim
\feh \lesssim -1.1$ and at $\feh \sim -0.5$. To further analyze the possible
cause of the deviation from linearity, we decided to add a third
dimension to the plot, by assigning to the plotted points a dimension
proportional to the mass of the corresponding clusters. By doing this, we
clearly observe that the scatter increases with the mass of the cluster.

An empirical way to tackle this issue is to select a sub-sample of clusters for
which an adequate linearity between RGB width and metallicity takes place.  A
robust choice is represented by the less massive clusters, which occupy the
bottom region of the observed trend. In particular, we selected all the
clusters with mass smaller than the median mass of the selected
sub-sample of GCs, namely Log\,(M/M$_{\odot}) = 5.22$. The
selected clusters \footnote{We highlight the fact that our sample of low-mass
GCs does not include any cluster with known metallicity spread
\citep{marino19a,marino15}.} have been marked as black-filled points
in the plot.  Then, we computed the weighted least-square linear relation
fitting the selected points, represented by the red dashed line in the plot,
whose expression has also been reported in the legend. The gray shaded area
surrounding the regression line represents the 95\% confidence interval, that
is the region having the 95\% probability to include the true regression line.
In order to verify the goodness-of-fit of the fit function, we computed the
adjusted determination coefficient R$_{adj}^2$, which is a measurement of the
ratio of variance between the residuals of the independent variable and
that of the dependent variable with respect to the regression line.
R$^2$ values range from 0, when the fit model fails to predict the data, to 1,
indicating a perfect fit. In our case, as reported in the legend, R$_{adj}^2 =
0.777$, thus indicating that the variance of the metallicity of the selected
sub-sample of clusters accounts for the $\sim 80\%$ of that of their RGB width.

The linear fit thus allows us to remove the effect of metallicity from the RGB
width of all the clusters in our database, by subtracting from their $W_{C
F336W,F438W,F814W}$, the residual with respect to the regression line. We will
refer to this new quantity as $\Delta W_{C F336W,F438W,F814W}$ or `normalized'
RGB width. We reported the corresponding values for each cluster in
Table~\ref{tab:ws}. Since the regression line has been determined empirically,
we did not take into account the error introduced by the linear fit relation.

\begin{figure*}
\centering
\includegraphics[width=\textwidth]{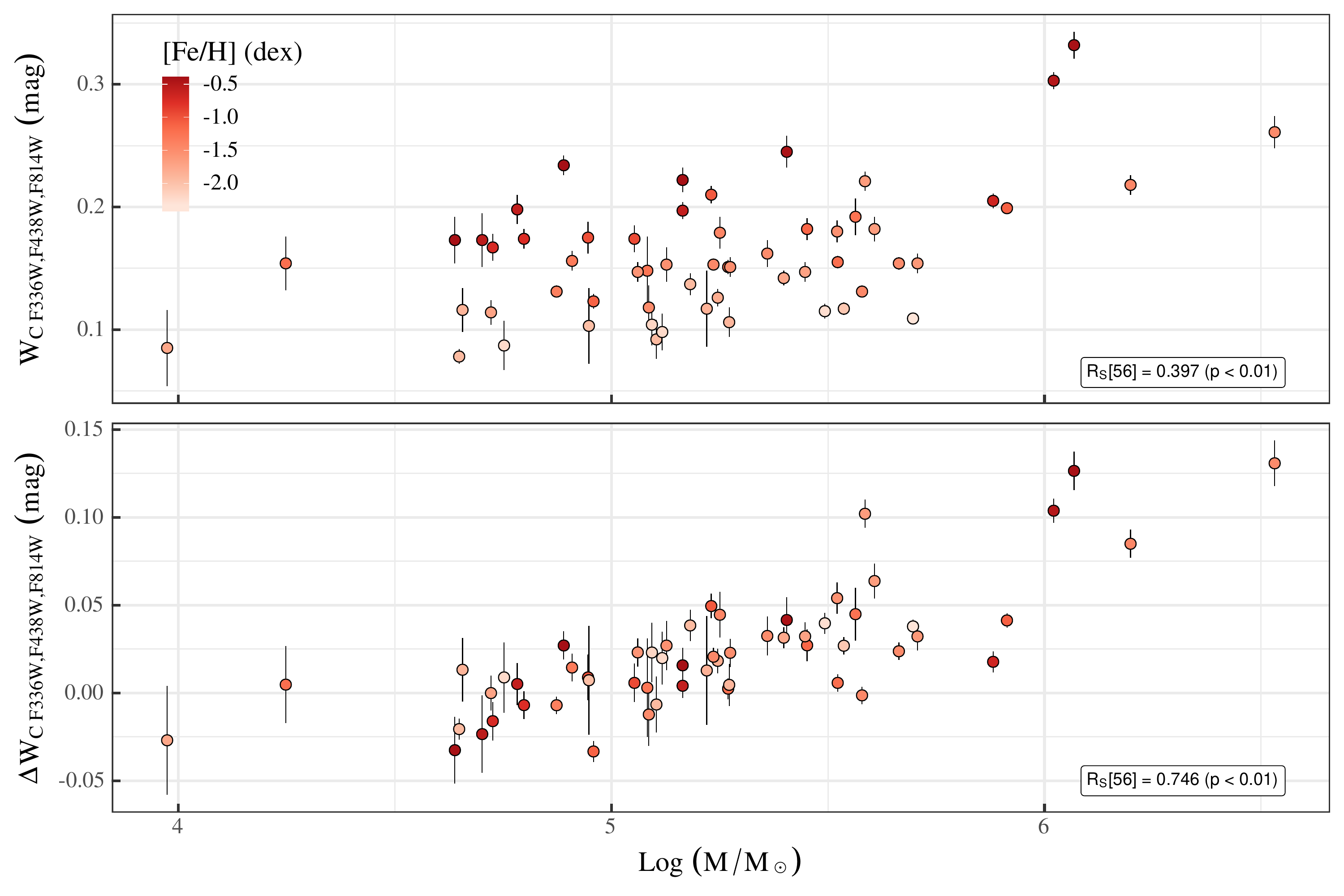} 
\caption{\textit{Top panel}: RGB width $W_{C F336W,F438W,F814W}$ vs.
	Log\,(M/M$_{\odot}$) for the Galactic GCs in our database. The color of
	each point corresponds to the metallicity of the cluster, according to
	the scale indicated in the legend. The bottom-right label reports the
	result of the Spearman's rank correlation test. \textit{Bottom panel}:
	as above but for the `normalized' RGB width $\Delta W_{C
	F336W,F438W,F814W}$.\label{fig:W_DW_Mass}} 
\end{figure*}

The effect of the inclusion of the metallicity variance in the RGB width is
clearly illustrated in the two panels of Figure~\ref{fig:W_DW_Mass}. In the top
panel, we plot $W_{C F336W,F438W,F814W}$ vs. Log\,(M/M$_{\odot}$) and assigned
to each point a color corresponding to the metallicity of the cluster,
according to the color scale shown in the legend. The result of the Spearman's
correlation test (R$_S[56] = 0.397$, p $ < 0.01$), has also been
indicated in the bottom right corner. The points appear aligned along
almost parallel directions, with the most metal-rich and the most metal-poor
clusters attaining, respectively, the highest and lowest values of the RGB
width, and the metal intermediate sitting between them, along the entire
outlined metallicity range. This plot shows, in a way reversed with respect to
Fig.~\ref{fig:lfit}, that the mass alone cannot explain the variance of the RGB
width but also that the metallicity has a clear predominant effect on the
observed trend characteristics. As shown in the bottom panel, once we subtract
the contribution of the metallicity given by the linear relation defined above,
we obtain a new trend, with all the points lined up along a single,
well-defined direction.  This new, monotonically increasing correlation, which
also appears linear over the entire metallicity range, suggests a strong
dependence from the mass of the properties of MPs, parameterized by the index
$\Delta W_{C F336W,F438W,F814W}$.

\section{Cluster mass and RGB width variation\label{sec:ncorr}}
In this section we use the Spearman's rank test to study the correlation
between the index $\Delta W_{C F336W,F438W,F814W}$ and all the cluster
parameters, except of course metallicity.  Since also \citet{milone17} computed
a `normalized' RGB width in the pseudo-color $C_{F275W,F336W,F438W}$, we
replaced, in this statistical batch, the formerly adopted index
$W_C$(\citetalias{milone17}) with their `normalized' index $DW_C$(\citetalias{milone17}).

\begin{figure*}
\centering
\includegraphics[width=\textwidth]{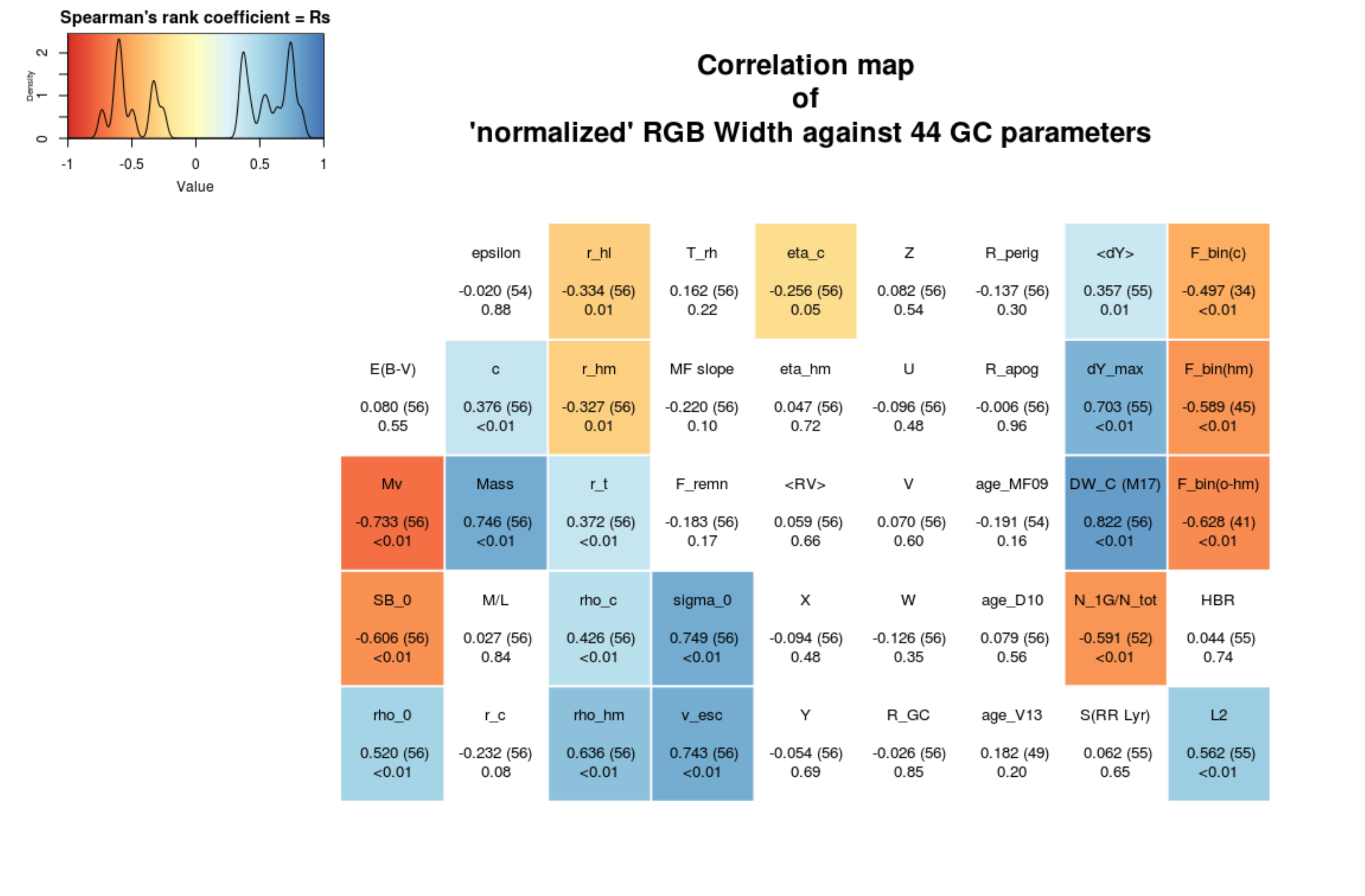}
\caption{As in Figure~\ref{fig:corrW_map} but for the `normalized' RGB
width. The cell relative to metallicity has been removed (see text for explanation).
\label{fig:corrDW_map}} 
\end{figure*}

As in the previous section, we show, in Figure~\ref{fig:corrDW_map}, the
correlation map of the results of the statistical test of the `normalized' RGB
width against the 44 GC parameters.

We observe now that a slightly larger fraction (21 out of 44) of parameters
score significant correlations. Moreover the kernel distribution shows that, at
odds with the previous case, the majority of them also have strong monotonic
correlations. The highest coefficient are found for $DW_C$(\citetalias{milone17})
(R$_S[56] = 0.822$, p $< 0.01$) and Mass (R$_S[56] = 0.746$, p $ < 0.01$).
Similarly to what found in the previous section, the first result confirms the
equal sensitivity of the `normalized' RGB index introduced here with the
corresponding quantity defined in \citet{milone17}. 

The second highest correlation value, relative to the cluster mass, is also 
corroborated by the highly significant, negative correlation existing with the total
cluster magnitude $M_V$ (R$_S[56] = -0.733$, p $ < 0.01$), being the luminosity
of a cluster a proxy of its total mass. Moreover, also other structural
parameters, directly linked to cluster mass content score high R$_S$ values, like
$\sigma_0$ (R$_S[56] = 0.749$, p $< 0.01$) and $\mathrm{v_{esc}}$ (R$_S[56] =
0.743$, p $ < 0.01$). In this context, we also notice the moderate/high
correlations with the parameters describing the fraction of binaries in
different regions of the cluster as well as with the ratio of 1G to the total
number of stars $N_{1G}/N_{tot}$.

Interestingly enough, some significant correlations present in the previous
statistical run disappeared, with some remarkable examples like the orbital
parameters R$_{GC}$ and R$_{apog}$, or
$\mathrm{age_{\,\text{\citetalias{dotter10}}}}$ and
$\mathrm{age_{\,\text{\citetalias{vandenberg13}}}}$. Among the new strong
correlations we highlight that with $\mathrm{\delta Y_{max}}$ (R$_S[56] =
0.703$, p $ < 0.01$).  Directly related to the latter is the moderate correlation
found for the color extension of L2, which is on first approximation dependent
on the internal helium spread of a cluster
\citep[e.g.][]{dantona02a,marino11,marino14,milone18b,tailo19a,tailo19}.

In conclusion, we observe that the RGB width $W_{C F336W,F438W,F814W}$ is a
sensitive indicator of the total internal chemical variations in Galactic GCs,
hence of their MP content.  The variance observed in $W_{C F336W,F438W,F814W}$
is mostly correlated with the cluster metallicity. After removing this
dependence, the `normalized' RGB width shows a significant, strong
correlation with the mass and the dynamic parameters linked to that, as well as
to maximum internal helium variations.

\section{General properties of the RGB width\label{sec:general}}
Any given chemical variation for stars at the base of the RGB, would result in
a different UV-optical color spread, for clusters with different metallicity
and age. Theoretical predictions of the variation of the index $W_{C
F336W,F438W,F814W}$ are, therefore, of primary importance in our
analysis, in order to properly compare the observations. Indeed, the whole
sample of GCs in our database spans a wide interval of metallicities ($-2.4
\lesssim $ \feh $\lesssim -0.4$) and ages ($\sim 2$ -- 14\,Gyr). To this aim,
we decided to analyze the effect of metallicity and age variations on the index
$W_{C F336W,F438W,F814W}$, by using appropriate theoretical models, as
explained in the following section.

\subsection{Trends from synthetic spectra analysis}
The RGB width $W_{C F336W,F438W,F814W}$ is strongly correlated to the internal
variation of helium and light elements, mostly through the CN, NH molecular
bands which in turn affect the ultraviolet and blue portion of the stellar spectrum
\citep[e.g.][]{marino08,sbordone11,milone13,monelli13}. Since the
absorption depth of a molecular band depends on the effective temperature and
gravity, we expect that for fixed variation of [C/Fe], [N/Fe], [O/Fe], and
helium mass content Y, stars with different atmospheric parameters would show 
different values of $W_{C F336W,F438W,F814W}$.

To investigate the effect of varying atmospheric parameters on the RGB width,
we took advantage of synthetic spectra to compute, across a grid of ages and
metallicities, the difference in the pseudo-color $C_{F336W,F438W,F814W}$
corresponding to a given variation of C, N, O and helium. The metallicity
grid-values are \feh $ = \mathrm{-2.45, -2.0, -1.5, -1.0, -0.5, 0.0}$\,dex,
while the age values are 2, 8, 14\,Gyr.  For each grid-point, we fetched two
isochrones, from the Dartmouth Stellar Evolution
Database~\footnote{\url{http://stellar.dartmouth.edu/models/}}
\citep[DSEP;][]{dotter07}, with the same \feh\ and age but different helium
content: one with canonical helium abundance (Y$\approx 0.25$) and the other
with helium enhanced by $\Delta Y = 0.045$ with respect to the canonical value.
The latter corresponds to the average internal helium variation derived in the
analysis of the maximum helium variation in 58 Galactic GCs, by
\citet{milone18b}. For each metallicity/age value, we estimated the \teff\ and
log\,g of the point 2.0\,mag brighter than the MSTO in F814W band,
along both the canonical and enhanced helium isochrone. Then, for each
point we generated, with ATLAS12 and SYNTHE codes
\citep{castelli05,kurucz05,sbordone07}, a synthetic spectrum with a
characteristic chemical abundance. For the point with the canonical helium we
simulated a spectrum with solar [C/Fe] and [N/Fe], and [O/Fe] = 0.3\,dex,
corresponding to the typical mixture of a 1G stars; for the point with enhanced
helium we simulated a spectrum  depleted in carbon and oxygen by 0.5 dex and
enhanced in nitrogen by 1.21\,dex, with respect to the previous one. This
abundance is indicative of the mixture of a 2G star. We integrated each
spectrum over the transmission curves of the WFC3/UVIS, ACS/WFC,
$\mathrm{U_{BESS}}$, $\mathrm{B_{BESS}}$,  $\mathrm{V_{BESS}}$ filters used in
this paper, to derive the synthetic $C_{F336W,F438W,F814W}$ and $C_{U,B,I}$
pseudo-colors. The absolute value of the difference between the 2G and 1G
pseudo-colors provides the theoretical measurement of the RGB width.

\begin{figure*}
\centering 
\includegraphics[width=\textwidth]{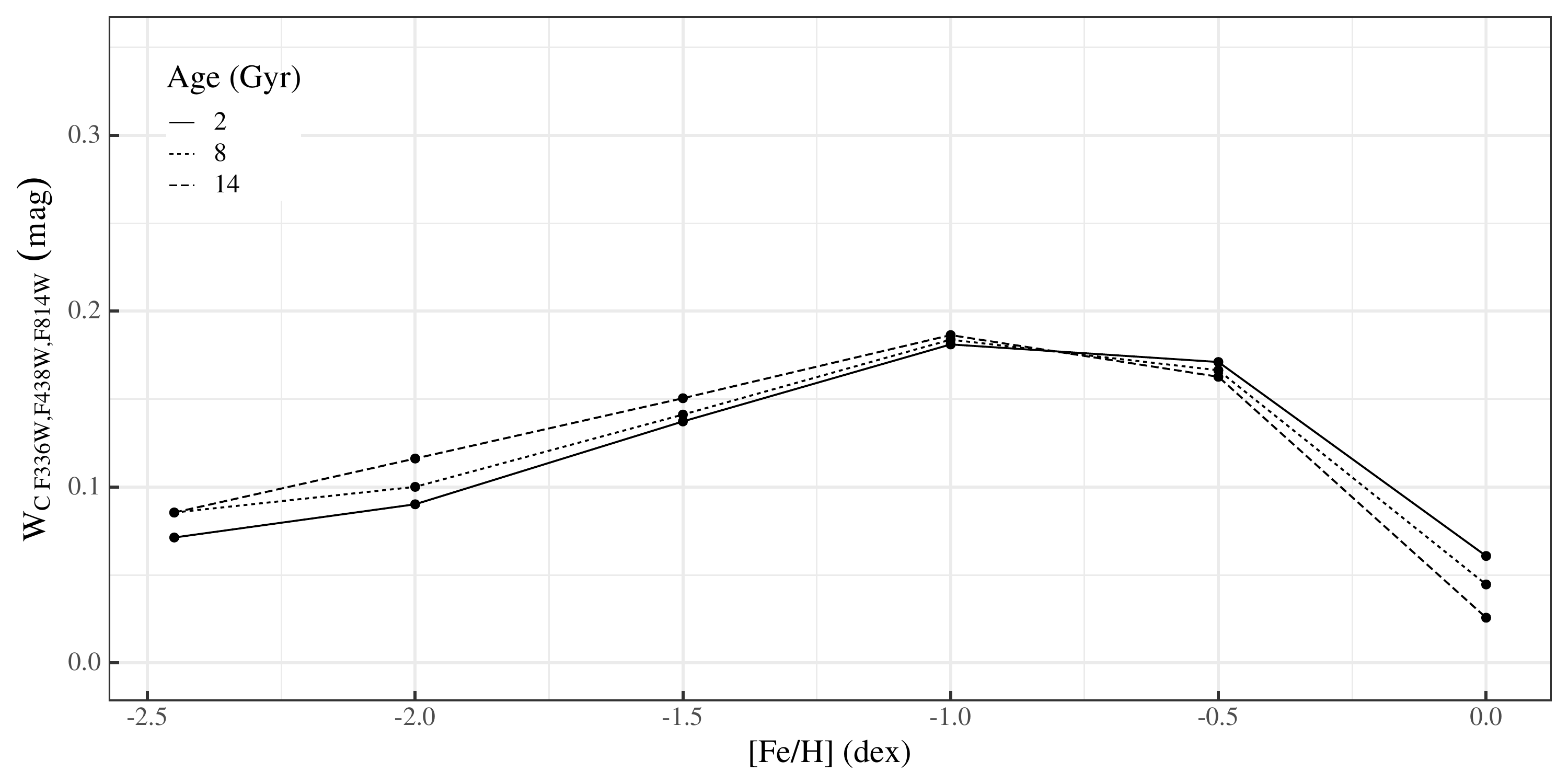}
\caption{Theoretical variation of the RGB width as a function of metallicity
\feh, for young-, intermediate-, and old-age models.\label{fig:age_trend}}
\end{figure*}

The trend of the theoretical $W_{C F336W,F438W,F814W}$ as a function of \feh, for
the three different grid values of age, is shown in Figure~\ref{fig:age_trend}.
It reveals that all the models  exhibit comparable trends, with the RGB width
almost linearly increasing from $\lesssim 0.1$ at \feh$=-2.45$ up to $\sim 0.2$
at \feh$=-1.0$, and decreasing first with a shallow slope at \feh$=-0.5$, and
then abruptly at solar \feh, where all the models reach their minimum value. We also
notice the that all the models attain similar values across the whole
metallicity range, with $W_{C F336W,F438W,F814W}$ lower for younger ages. The
profile is inverted at \feh$ \gtrsim -1.0$, where $W_{C F336W,F438W,F814W}$
starts to be higher for younger ages. Finally we observe the largest scatter at
solar metallicity with a difference between the models of $\approx 0.03$\,mag.
We can conclude that we can safely perform a direct comparison among the
observed RGB width of young-, intermediate- and old-age clusters.

As a final remark, we notice that the values of $C_{F335W,F438W,F814W}$ and
$C_{U,B,I}$ are almost the same within 0.02 mag, with the $\mathrm{U_{HIGH}}$,
$\mathrm{B_{BESS}}$ and $\mathrm{I_{BESS}}$, used in this paper for Terzan\,7,
providing wider RGB spread.  This fact demonstrates that we can directly
compare Terzan\,7, for which only ground-based photometry is available, with
the other clusters.

\section{Comparison between Galactic and extragalactic clusters\label{sec:compar}}
So far, we analyzed the properties of the RGB width of the Galactic
GCs included in our database. Thanks to the availability of
homogeneous measurement for a large dataset of parameters, this
selection constitutes the best reference system for the comparison of the
observed properties of stellar systems presenting peculiar characteristics. In
this respect, the GCs \ngc2419 and Rup\,106 are remarkable examples.
Indeed, the first is the most external cluster of the Milky Way
\citep[R$_{GC} = 90.57$\,kpc;][]{baumgardt19}, while the second
belongs to a peculiar class of clusters not hosting MPs. Moreover, the recent
discovery of MPs in MC clusters, makes them the ideal target for a comparative
study of the MP properties in stellar systems belonging to different
galaxies.

\begin{figure*}
\centering
\includegraphics[width=\textwidth]{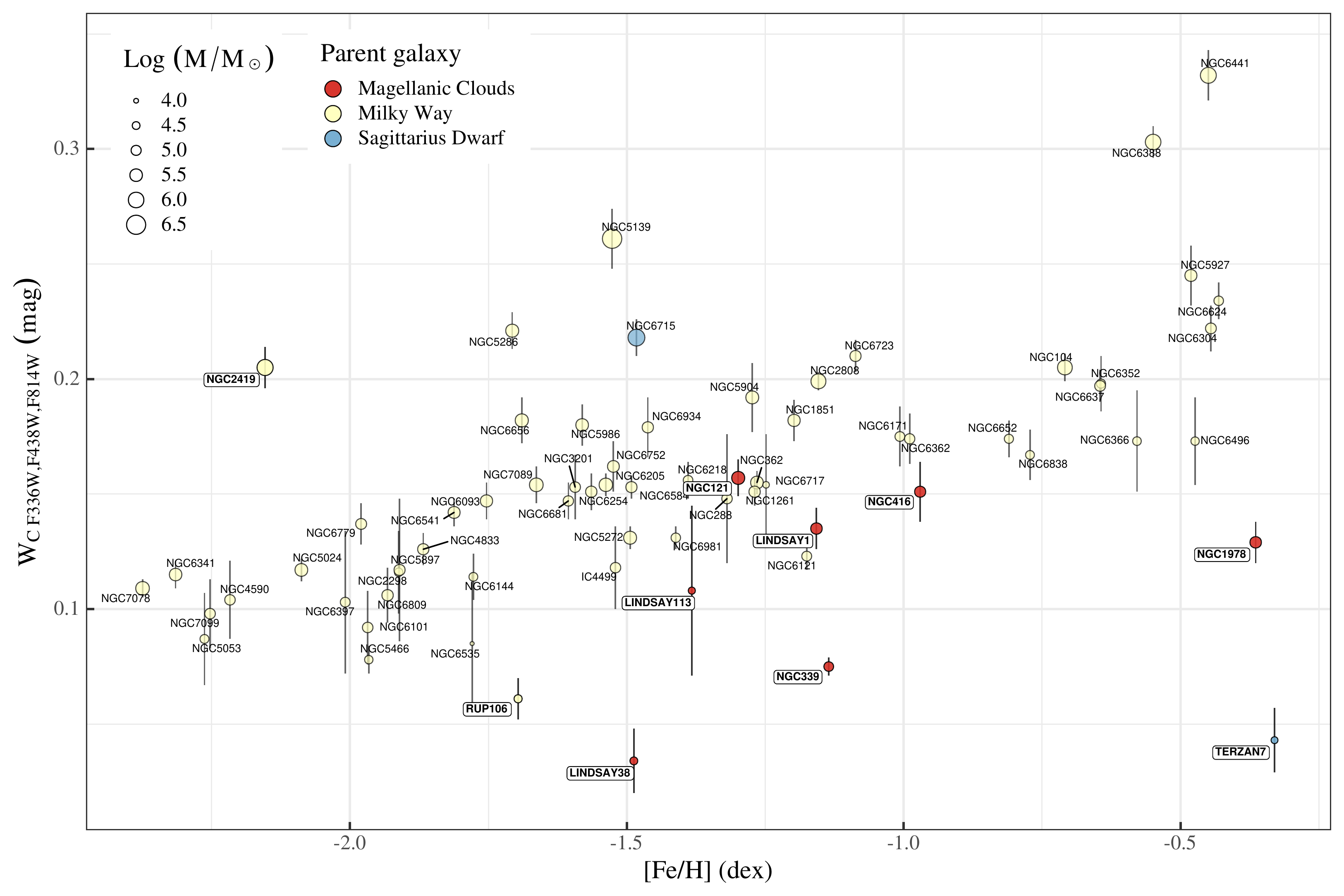}
\caption{$W_{C F336W,F438W,F814W}$ vs. \feh\ for all the analyzed GCs. The color
	and dimension of each point refers to the parent galaxy and the mass of
	the corresponding cluster. For the sake of clarity, the most remarkable
	Galactic GC outliers, namely \ngc2419 and Rup\,106, as well as the
	Magellanic Cloud clusters and Terzan\,7 have been indicated with a
	boxed label.\label{fig:W_vs_FeH}} 
\end{figure*}

We decided, therefore, to reproduce the plot of Fig.~\ref{fig:lfit} by
including these new objects. The new plot is shown in
Figure~\ref{fig:W_vs_FeH}, where each point has been labeled with the name of
the corresponding cluster, and colored according to its parent galaxy. As
already mentioned in Sect.~\ref{sec:corr}, since \ngc6715 is the core of the
Sagittarius dwarf, it has been colored azure like Terzan\,7, while
Milky Way and Magellanic Cloud GCs have been represented as light yellow and
red points, respectively. We see in the plot that, among the Milky Way GCs,
\ngc2419 represents a remarkable outlier. This cluster, indeed, which has been
found to host at least four different stellar populations with extreme helium
abundances, as recently reported in \citet{zennaro19} and \citet{larsen19},
displays a RGB width almost a factor of two higher than that typical of MW GCs
at low metallicity regime.  On the other side, Rup\,106 has an unusually small
RGB width for its metallicity.  The location of Rup\,106 in the plot reflects
the unusual property of this globular which is one of the few MW clusters
hosting a single stellar population, as reported in \citet{villanova13}
and confirmed by \citet{dotter18}.

Another interesting feature visible in this plot is the location of the
extragalactic clusters, which have, on average, RGB width values lower that of
MW GCs with comparable mass and metallicity. In particular, only three MC
clusters seem to follow the general trend observed for the MW GCs, namely
\ngc121, \ngc416 and Lindsay\,1, while \ngc1978, Lindsay\,113, \ngc339,
Terzan\,7, and Lindsay\,38, attain $W_{C F336W,F438W,F814W}$ values lower than
those typical of GCs. We also notice that the RGB width of Lindsay\,38 and
Terzan\,7 is smaller than that of Rup\,106, thus suggesting that both the
clusters are likely consistent with a single stellar population. This represent
a new finding for Terzan\,7 while it is consistent with that obtained by
\citet{martocchia19} and \citet{li19} for Lindsay\,38 and
Lindsay\,113, respectively.

Despite the typical lower RGB width values, the overall trend of the
extragalactic GCs seems consistent with that of the Galactic GCs, with RGB
width values typically higher for the most metal-rich clusters and at a given
metallicity, higher for the most massive clusters.  As a consequence, we
applied to the new sample, the procedure to remove from the observed RGB width
the effect of the metallicity.

\begin{figure*}
\centering
\includegraphics[width=\textwidth]{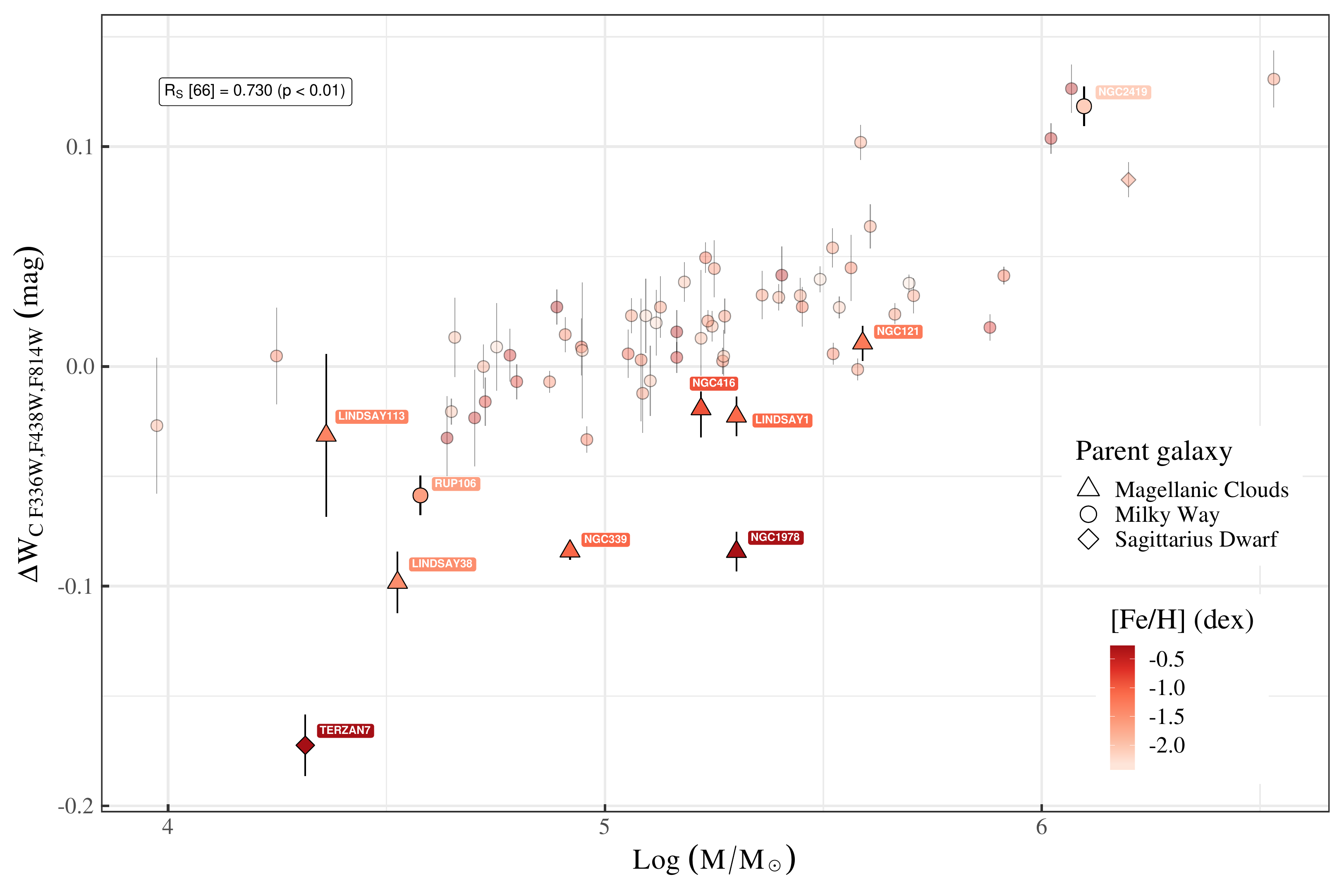}
\caption{$\Delta W_{C F336W,F438W,F814W}$ vs. Log\,(M/M$_{\odot})$  for all the
	analyzed GCs. The color and shape of each point refers, respectively,
	to the metallicity and the parent galaxy of the relative cluster. For
	the sake of comparison, the Galactic GCs \ngc2419 and Rup\,106 and the
	extragalactic GCs have been also labeled. We notice the systematically lower
	location of the extragalactic globulars with respect to the bulk of MW
	clusters.\label{fig:DW_vs_Mass}}
\end{figure*}

The new plot is shown Figure~\ref{fig:DW_vs_Mass}, where similarly to
Fig.~\ref{fig:W_DW_Mass}, we used a color scale to map the metallicity of the
clusters. In this case, in order to distinguish the parent galaxy of each
globular, we used a different shape for the plotted points. The result of the
Spearman's correlation rank test (R$_S[66] = 0.730$, p $ < 0.01$) indicates a
significant, strong monotonic relation between the `normalized' RGB width and
the mass of the clusters, similarly to what found for the Galactic GC
subsample. The plot, however, also reveals that the extragalactic
clusters systematically deviate from the general behavior of the MW GCs. We
observe, indeed, that while \ngc2419 and Rup\,106 follow the general trend of
the other Galactic GCs, all the extragalactic GCs are located on the faint side
of the bulk of MW clusters, with the exception of Lindsay\,113, although no
firm conclusion is possible for this cluster given its large error bar.  As
expected, with the exception of \ngc121, \ngc416 and Lindsay\,1, all the other
extragalactic GCs show a significant scatter with respect to the main trend,
with Terzan\,7 attaining the lowest $\Delta W_{C F336W,F438W,F814W}$ value.

We conclude that the lower intrinsic color spread of the RGB width, observed
for the MC clusters as well as for Terzan\,7 seems to be an inherent properties
of these clusters. These results seem consistent with a scenario where MPs in
both Galactic and MC GCs follow the same relation with the mass
of the host cluster, but the ancient Galactic GCs have lost a significant
amount of their mass. Moreover, this is consistent with the result of
\citet{zennaro19}, who found that GCs with large perigalacticon host, on
average, larger fractions of 1G stars than the remaining Galactic GCs, thus
indicating that the interaction with the Milky Way affects the properties of
MPs in GCs. As an alternative, the properties of MPs in the ancient GCs, formed
in the Milky Way at high redshifts, and in the extragalactic clusters could
eventually depend on the different environment at the epoch of formation.

\section{RGB width and cluster age\label{sec:age}}
\begin{figure*}
\centering
\includegraphics[width=\textwidth]{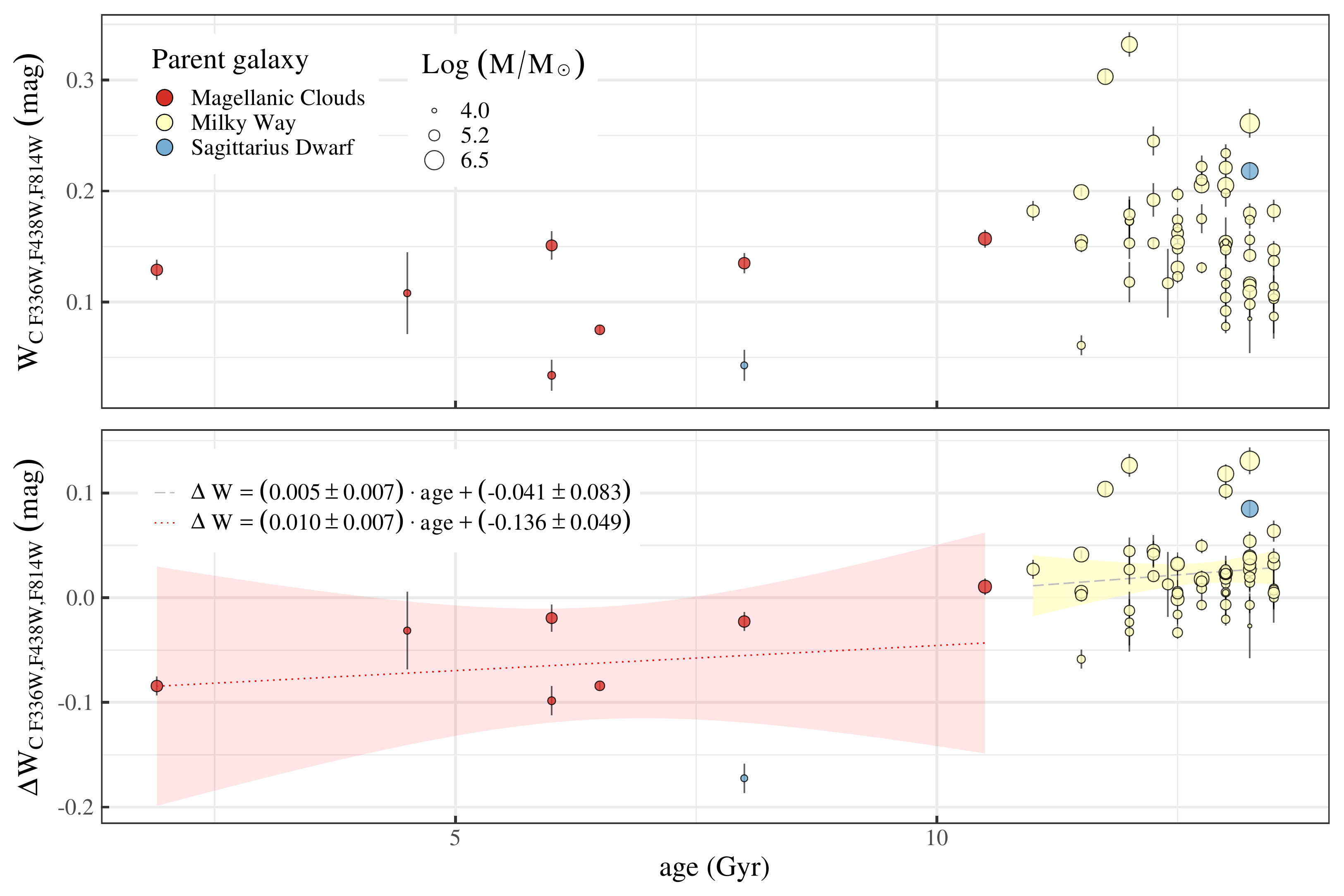}
\caption{\textit{Top panel}: $W_{C F336W,F438W,F814W}$ as a function of the
	cluster age.  As in Figure~\ref{fig:W_vs_FeH}, the dimension of each
	point is proportional to the corresponding cluster mass, while the
	color refers to the parent galaxy.\textit{Bottom panel}: $\Delta W_{C
	F336W,F438W,F814W}$ vs.  age. The dashed gray and red dotted regression
	line fits, respectively, the Galactic sample (including \ngc6715) and
	the extragalactic clusters. The yellow and red shaded area
	represents the 95\% confidence interval of the fit relationships, for
	which we also reported the corresponding
	equations.\label{fig:W_DW_vs_age}}
\end{figure*}

The discovery, confirmed by both spectroscopic and photometric analyses, of
internal N and He variations in MC globulars older than $\sim 2$\,Gyr, obtained
in the past few years
\citep[e.g.][]{chantereau19,lagioia19,hollyhead17,martocchia18a,niederhofer17a,niederhofer17},
together with the lack of detectable spread in younger clusters, raised new
questions on the possible effect of the age as a driver for the appearance of
the MPs in GCs \citep[see][]{martocchia19}. We decided to investigate the
relation between RGB width and age by taking into account the whole sample of
clusters in our database, which spans an age range of about 11\,Gyr. The top panel of Figure
~\ref{fig:W_DW_vs_age} displays $W_{C F336W,F438W,F814W}$ as a function of the cluster
age. We used the age values from \citet{dotter10,dotter11} and \citep{milone14}
for the Galactic GCs and Terzan\,7, while for \ngc121, \ngc339, \ngc416 and
Lindsay\,1 we considered the values provided in \citet[][and references
therein]{lagioia19}. The ages of Lindsay\,38, Lindsay\,113 have
been taken from \citet{martocchia19}, while that of
\ngc1978 from \citet{mucciarelli07}. We observe that the extragalactic
clusters, that are, with the exception of \ngc121 (age
$\sim$ 10.5\,Gyr), all younger than the MW GCs, have RGB width values
similar to those of the oldest Galactic GCs. The recent conclusion by
\citet{martocchia19}, that the nitrogen content in GCs is mostly correlated with the
cluster age, seems to be not consistent with the present result. We notice,
however, that these authors based their analysis on the standard deviation of
the color distribution of the RGB stars in a different color combination,
namely $C_{F343N,F438W,F814W}$, which is similar to $C_{F336W,F438W,F814W}$ but
exploits the narrow band F343N filter instead of F336W, and
$C_{F336W,F438W,F343N}$, which is given by the sum of the colors
$m_{F336W}-m_{F438W}$ and $m_{F343N}-m_{F438W}$. For this reason they have
a much smaller Galactic GC sample, composed by the three clusters \ngc104
(47\,Tuc) \ngc2419 and \ngc7078 (M\,15).

As demonstrated in Sect.~\ref{sec:corr} and \ref{sec:general}, a fixed
variation of [C/Fe], [N/Fe], [O/Fe] and Y in clusters with different
metallicities results in different RGB widths \citep[see also][]{lagioia18,
milone18b}. As a consequence, for a proper comparison between the nitrogen
abundance variation and any other GC parameter, we must take into account the
effect of the metallicity.  The bottom panel of Figure~\ref{fig:W_DW_vs_age}
displays $\Delta W_{C F336W,F438W,F814W}$ as a function of the cluster age.
In this plot we see that the location of the extragalactic
clusters along the vertical axis is somewhat lower than that of the Galactic
GCs, although several old Galactic GCs share similar $\Delta W_{C
F336W,F438W,F814W}$ values as intermediate-age MC clusters. We also notice that
there is a dependence of the RGB width from the cluster mass, with the most
massive GCs, attaining higher $\Delta W_{C F336W,F438W,F814W}$ values.
Unfortunately, the possibility to include the mass variance into the index
$\Delta W_{C F336W,F438W,F814W}$ is hampered  by the lack of precise
knowledge of the initial mass of the clusters.  In fact, the values of the
initial mass, available only for the Galactic GCs (H. Baumgardt, private
communication), have been computed under the hypothesis of constant Galactic
potential and unperturbed orbits.  Dynamical evolution in the galactic
potential, though, dramatically changes the pristine mass of a GC.

As a matter of fact, the heterogeneous physical conditions of the MW and its
satellites might have significantly altered the evolutionary paths of their
respective globulars. This scenario is also complicated by the fact that a
substantial portion of the GCs in the MCs, for instance, was formed at
different redshifts and as such, evolved with different boundary conditions.
We can speculate whether in our case a different dependence of the index
$\Delta W_{C F336W,F438W,F814W}$ from the cluster age in the two subsystems,
namely the MW and the MC GCs, can provide some hints about the role of the
environment on the general properties of the MPs in GCs. On this purpose we
computed the weighted least-square linear relations fitting the two different
GC groups, with the Galactic group including the Sagittarius Dwarf cluster
\ngc6715 for internal consistency with our analysis. The dashed gray and dotted
red lines in the plot represent the two regression lines, for which we
also reported the corresponding equations in the legend, and the yellow and
red shaded areas the respective 95\% confidence intervals. 

The errors indicate that the slope and the intercept of the two regression
lines are consistent within 1\,$\sigma$. In order to estimate the significance
of the difference between the two fitting relations we run the
analysis-of-variance (ANOVA) test \citep{chambers92}. The first part of the
test is based on the assumption of interaction between the predictor variable
``age'' and the categorical variable ``parent galaxy''; the second relies on
the assumption of no interaction between the same two variables. In the first
case we found that ``age'' has a significant effect (p $ < 0.01$) on the
response variable ``$\Delta W$'', while the effect of ``parent galaxy'' is not
significant (p = 0.152).  Also, the interaction between ``age'' and ``parent
galaxy'' is insignificant (p = 0.559). In the second test, we obtained the same
outcome for the significance of the effect of the predictor (p-value $< 0.01$)
and categorical (p-value = 0.150) variable. We conclude that if any relation
exists between age and `normalized' RGB width it seems to be independent of the
galaxy under consideration.

\section{Summary} \label{sec:summary}
The evidence of internal light-element variations in several GCs of the
Magellanic Clouds and Fornax
\citep[e.g.][]{lagioia19,martocchia18a,niederhofer17,larsen14a} indicates that
MPs are common features of extragalactic GCs.  

The wealth of observations of MPs in GCs demands a common metrics to directly
compare the various clusters and to constrain the physical parameters that
determine the MP phenomenon.  To this aim, we undertook an extensive study of
the MPs in 68 Galactic and extragalactic GCs, by measuring the color extension
of the stars at the base of the RGB, in the UV-optical pseudo-color
$C_{F336W,F438W,F814W}$, which is sensitive to the stellar content of C, N, O
and helium \citep{marino08}, and is available for a large number of archival
observations. Images through these three filters can be collected from
ground-based facilities also in the post-\hst\ era, thus allowing the extension
of the present analysis to distant GCs. 

The analysis of the monotonic trend, performed with a series of Spearman's rank
correlation tests, of the RGB width, $W_{C F336W,F438W,F814W}$, of 58 Galactic
GCs observed in F275W \citep{milone17}, against 45 different GC observational,
structural and morphological GC parameters, shows that the RGB width is mostly
dependent on the cluster metallicity (see Fig.~\ref{fig:corrW_map}). We also
find an almost perfect monotonic correlation between the $W_{C
F336W,F438W,F814W}$ and the similar quantity $W_{C F275W,F336W,F438W}$ by
\citet{milone17}.  This result demonstrates the high effectiveness/cost ratio
of the filter choice adopted in this work for the measurement of the extension
of the MPs. Indeed, observations in F275W are very expensive in terms of
telescope time and can be done by {\it HST} alone.  By empirically removing the
dependence of $W_{C F336W,F438W,F814W}$ from \feh, we defined a
metallicity-free RGB width, $\Delta W_{C F336W,F438W,F814W}$, that we dubbed
`normalized' RGB width, which shows a significant monotonic correlation with
the cluster total mass (see Fig.~\ref{fig:W_DW_Mass}). The Spearman's
rank correlation test between $\Delta W_{C F336W,F438W,F814W}$ and the global
GC parameters reveals the highest degree of monotonic correlation with the
cluster mass, thus indicating the prominent role of this parameter in
determining the complexity of the MP phenomenon in the Galactic GCs.  The
`normalized' RGB width also shows highly significant correlation coefficients
with the core velocity dispersion, $\sigma_0$ and the escape velocity,
$v_{esc}$. A mild correlation with the fraction of 1G stars $N_{1G}/N_{tot}$ is
also present. Our findings are compatible with the high correlation found by
\citet{baumgardt19} between $N_{1G}/N_{tot}$ and $v_{esc}$ for the sample of
Galactic GCs studied in \citet{milone17}.

We compared the properties of the bulk of MW GCs and a set of Galactic and
extragalactic GC,  including the distant MW globular \ngc2419 and Rup\,106,
which is one of the few MW GCs found to host a single stellar population
\citep[see][]{villanova13,dotter18}, and seven MC clusters plus Terzan\,7,
associated with the stream of the Sagittarius Dwarf \citep{sbordone05}. The
location of the new clusters in the $W_{C F336W,F438W,F814W}$ vs. \feh\ diagram
shows that they are evident outliers, with the remarkable cases of the young-
and intermediate-age MC clusters and Terzan\,7, which attain systematically
lower RGB width values with respect to the MW GCs. Furthermore, for the latter
cluster we observe a RGB width smaller than that of Rup\,106, which is a
Galactic GC with a single stellar population. We therefore conclude that
Terzan\,7 likely is a single population cluster.

The comparison between the `normalized' RGB width of the MW and extragalactic
GCs as a function of the cluster mass demonstrates that the extragalactic GCs
systematically deviates from the bulk trend of the Galactic globulars, with the
extragalactic systems attaining lower values of $\Delta W_{C
F336W,F438W,F814W}$.  This fact suggests that the Magellanic Cloud GCs exhibit
smaller internal light-element variations than Galactic GCs with similar
present-day masses. 

Arguably, the observed difference depends on the different physical conditions
in Galactic and extragalactic proto-GCs at the epoch of formation. As an
alternative, we speculate that the MPs in MW and extragalactic GCs follow a
unique trend with the initial cluster mass, but ancient Galactic GCs lost a
significant amount of their mass. 

In a recent paper, \citet{martocchia19} measured the standard deviation of the
RGB of eight Magellanic Cloud clusters with ages between $\sim$ 2 and
$\sim$11 Gyr and that of the massive Galactic GCs M\,15, 47\,Tuc and
NGC\,2419,, that are $\sim 12$ - 13 Gyr old.  
They used two pseudo-colors that are mostly sensitive to nitrogen
abundance, namely the index $C_{F343N,F438W,F814W}$ and
$C_{F336W,F438W,F343N}$. 
  
Based on the correlation between the standard deviation of the RGB and the
cluster age, they conclude that the internal abundance variation
strongly depends on the cluster age. We find no evidence for correlation
between the RGB width in $C_{F336W,F438W,F814W}$ and the
cluster age, although Magellanic Cloud clusters with ages between $\sim 2$ and
11\,Gyr exhibit, on average, lower values of $W_{C F336W,F438W,F814W}$
and $\Delta W_{C F336W,F438W,F814W}$ than for the Galactic GCs.
However, the evidence that the RGB width mostly depends on cluster metallicity
and mass, together with the fact that it is not possible to directly compare
present day masses of Galactic and extragalactic clusters with different ages,
prevents us from any conclusion on the possibility that the cluster age affects
the RGB width through internal nitrogen variation.

\facilities{HST (WFC3); VLT (FORS2)}

\software{img2xym (Anderson et al. 2006), EsoReflex (Freudling et al. 2013),
DAOPHOT and ALLSTAR (Stetson 1987, 1994), DAOMASTER (Stetson 1990), ALLFRAME
(Stetson 1994), ATLAS12 (Castelli 2005; Kurucz 2005), SYNTHE (Sbordone et al.
2007)}

\acknowledgments
This work has received funding from the European Research Council
(ERC) under the European Union's Horizon 2020 research innovation
programme (Grant Agreement ERC-StG 2016, No 716082 `GALFOR', PI:
Milone), and the European Union's Horizon 2020 research and innovation
programme under the Marie Sklodowska-Curie (Grant Agreement No 797100,
beneficiary: Marino). APM and MT acknowledge support from MIUR through the
the FARE project R164RM93XW `SEMPLICE'.

\bibliography{ms}

\end{document}